\documentclass[12pt,oneside]{article}
\usepackage{setspace}
\usepackage{amsmath, amsthm, amssymb}
\usepackage{verbatim, appendix}

\usepackage{graphicx}
\usepackage{caption}
\usepackage{enumitem}

\usepackage{times}
\usepackage{amsmath}
\usepackage{fullpage}
\usepackage{bm}
\usepackage{bbm}
\usepackage[normalem]{ulem}
\usepackage{amsfonts}
\usepackage{mathrsfs}
\usepackage{tikz} 
\usetikzlibrary{arrows.meta, positioning,shapes.geometric}

\newcommand{\stkout}[1]{\ifmmode\text{\sout{\ensuremath{#1}}}\else\sout{#1}\fi}
\usepackage{amsthm} 
\newtheorem{definition}{Definition}[section]
\newtheorem{proposition}{Proposition}[section]
\usepackage{url}

\usepackage{amsmath}
\usepackage{algorithm}
\usepackage{xurl}
\usepackage[hidelinks]{hyperref}
\makeatletter
\def\BState{\State\hskip-\ALG@thistlm}
\makeatother

\usepackage[top=1in, bottom=1.0in, left=1.0in, right=1.0in]{geometry} 

\begin{document}
\normalem 
\pagenumbering{gobble}
\begin{titlepage}
    \centering
    \vspace*{\fill}

    {\LARGE\bfseries
    Inherited Local Shrinkage for Adaptive Functional Form Discovery
    \par}

    \vspace{1.5cm}

    {\large
    Jacob C. Schultz\\
    Daniel Taylor-Rodriguez
    \par}

    \vspace{0.5cm}

    {\normalsize
    Fariborz Maseeh Department of Mathematics + Statistics\\
    Portland State University
    \par}
    
    \vspace*{\fill}
\end{titlepage}

\pagenumbering{arabic}

\begin{abstract}
\hyphenpenalty=10000
\exhyphenpenalty=10000

Statistical models frequently employ nonlinear effects, interactions, and multiresolution basis expansions to represent complex relationships between predictors and responses. Determining the appropriate level of model complexity, however, remains a fundamental challenge. This paper introduces \textit{Inherited Local Shrinkage} (ILS), a Bayesian global-local shrinkage framework for adaptive functional form discovery that incorporates strong heredity among hierarchically related predictors.

ILS propagates shrinkage through directed acyclic graphs (DAGs) encoding relationships among predictors, allowing increasingly complex terms to contribute only when supported by their simpler ancestors. This construction provides a common prior framework for a broad class of hierarchical regression models. We demonstrate its application to polynomial regression, as well as multiresolution modeling with two-dimensional Haar wavelet and eigenvalue-ordered spline (EVOS) bases. Although these examples illustrate specific basis systems, ILS is not tied to any particular basis construction and can be applied to any regression model in which relationships among predictors can be represented by a DAG, regardless of the graph's size or topology.

Simulation studies demonstrate that ILS achieves competitive predictive performance relative to existing heredity-constrained methods while recovering underlying functional structure. Spatial applications further illustrate its ability to adaptively regulate model complexity across multiple resolutions, identifying where finer-scale detail is supported by the data. Together, these results establish ILS as a flexible framework for hierarchical complexity control and adaptive functional form discovery in Bayesian regression.
\end{abstract}

\section{Introduction}
\label{ch:intro}

Identifying complex relationships between predictors and a response is a central objective in regression modeling. In many scientific applications, additive linear effects are insufficient to characterize the underlying data-generating process. Nonlinear trends, interactions among predictors, and multiscale spatial or functional structure often require richer representations constructed from basis expansions such as polynomial terms, splines, wavelets, or multi-resolution spatial bases \cite{friedman1991mars,wood2017generalized,Daubechies1992,Mallat1999}. While these expansions substantially increase modeling flexibility, they also generate large collections of structured predictors whose complexity must be adaptively controlled.

This paper introduces inherited local shrinkage (ILS), a Bayesian global-local shrinkage framework for adaptive functional form discovery in structured regression models. The proposed method propagates shrinkage through a directed acyclic graph (DAG) encoding hierarchical relationships among predictors, ensuring that increasingly complex terms receive support only when simpler parent structure is retained. By embedding strong heredity directly into the local shrinkage mechanism, ILS adaptively suppresses unsupported high-order terms while preserving interpretable and well-formulated models.

Flexible statistical modeling strategies based on basis expansions are widely used to capture nonlinear, interaction, and multiscale structure in regression problems. Polynomial bases, spline expansions, wavelets, and other basis functions provide increasingly expressive representations capable of approximating complex functional relationships while retaining computational and interpretive structure \cite{wood2017generalized,trefethen2013approximation,Daubechies1992,Mallat1999}. Increasing model flexibility, however, introduces an important methodological challenge: the number of candidate predictors may grow rapidly as interactions, higher-order terms, or fine-resolution basis functions are added. For example, in polynomial regression, the number of interaction terms grows combinatorially with model order.

Determining the appropriate level of model complexity is therefore fundamental. Overly restrictive models may fail to capture meaningful nonlinear structure, whereas excessively flexible models may overfit the data, inflate variance, reduce interpretability, and/or degrade predictive performance \cite{breiman1995garrote,zou2006adaptive,bach2012structured}. This tradeoff is particularly pronounced in modern regression settings where large collections of potentially redundant basis functions are introduced to ensure sufficient expressive capacity.

Shrinkage and regularization techniques provide a principled framework for this problem by adaptively suppressing unsupported coefficients while preserving significant signal. Classical frequentist methods, including ridge regression, the lasso, and the elastic net, implement this via penalized optimization \cite{hoerl1970ridge,tibshirani1996lasso,zou2005elastic}. In contrast, Bayesian frameworks typically induce shrinkage through prior distributions on the regression coefficients, offering additional benefits such as posterior uncertainty quantification and information sharing across parameters \cite{george1993variable,kyung2010bayesian}.

Both paradigms utilize \textit{global-local} structures to facilitate sparse modeling. In frequentist contexts, models such as the adaptive Lasso utilize coefficient-specific weights alongside global regularization hyperparameters \cite{zou2006adaptive,zhao2009composite}. Bayesian global-local shrinkage models decouple the general level of shrinkage (global) from the specific behavior of each coefficient (local) by utilizing hierarchical variance components. This allows for a heavy-tailed local component that permits large coefficients to remain largely unaffected while aggressively shrinking small, noise-dominated effects toward zero. The horseshoe prior is a prototypical example of this structure, functioning as a common default for sparse regression due to its ability to preserve significant signals while effectively suppressing noise \cite{Bhadra2017}.

In many structured regression settings, predictors are not independent entities but instead possess explicit hierarchical relationships. For example, a quadratic term depends conceptually on its associated linear term, and a two-way interaction depends on both of its constituent main effects. Similar relationships arise in wavelet bases, multiresolution spatial bases, and functional expansions where fine-scale basis functions refine coarser structures. In such settings, it is desirable for increasingly complex terms to enter the model only when supported by simpler parent structure.

This principle can be formalized through heredity constraints. Under strong heredity, a higher-order term may only enter the model if all associated lower-order parent terms are also included. Strong heredity has long been recognized as important for interpretability and model invariance \cite{nelder1977logical,hamada1992analysis,peixoto1987hierarchical}. For example, a model containing an interaction term without its associated main effects can produce coefficients whose interpretations depend on arbitrary coding transformations of the predictors. Peixoto demonstrated that polynomial regression models are invariant to coding transformations if and only if they satisfy strong heredity conditions \cite{Peixoto1990}. Strong heredity has also been shown to possess favorable asymptotic selection properties. Under strong heredity constraints, posterior probability concentrates on a unique smallest well-formulated model nesting the truth, whereas weak heredity may asymptotically yield multiple competing minimal models when the true model itself is not well formulated \cite{taylorrodriguez2015heredity}.

A substantial literature has developed methods for enforcing heredity constraints in regression. Frequentist approaches include hierarchical lasso formulations and structured sparsity penalties \cite{bien2013lasso,jenatton2011structured,bach2012structured}. Bayesian approaches have similarly incorporated heredity through constrained model spaces or hierarchical variance specifications \cite{Chipman1996,taylorrodriguez2015heredity,griffin2017hierarchical}. While effective in low-order polynomial and interaction settings, many existing methods are tailored to relatively shallow hierarchical structures. Extending them to deeper polynomial expansions, irregular inheritance structures, or multiscale basis systems often requires increasingly elaborate grouping constraints, custom prior specifications, or level-specific tuning parameters.

ILS addresses these limitations by enforcing strong heredity through constraints on local shrinkage parameters within a global-local prior structure. Rather than requiring separate prior constructions for different hierarchy depths or interaction orders, the method propagates shrinkage through an arbitrary directed acyclic graph representing inheritance relationships among predictors. Under this construction, child coefficients receive at least as much shrinkage as their parents, ensuring that unsupported lower-order structure automatically suppresses increasingly complex descendants. Consequently, the framework adaptively determines the level of nonlinear, interaction, or multiscale complexity warranted by the data while preserving well-formulated hierarchical structure.

The proposed framework is intentionally general and applies beyond polynomial regression. Any regression basis admitting a DAG representation may be accommodated, including interaction models, multiscale spatial bases, spline hierarchies, and wavelet bases. This allows ILS to unify heredity-constrained shrinkage across a broad class of structured regression problems while preserving a comparatively simple prior specification.

The remainder of the paper is organized as follows. Section 2 introduces the notation and background necessary for the proposed framework, including polynomial basis expansions, inheritance DAGs, and related heredity-constrained regression methods. Section 3 presents the inherited local shrinkage methodology, establishes the associated heredity properties, and describes posterior computation and implementation details. Section 4 evaluates the proposed method through simulation studies comparing ILS to existing hierarchical shrinkage and regularization approaches under varying inheritance structures. Section 5 demonstrates the methodology in both time series and 2-dimensional spatial modeling applications. Finally, Section 6 synthesizes the results and discusses the broader role of ILS in adaptive functional form discovery.
\section{Background and Notation}

Consider the following linear regression model in $k$th-order polynomial space
\begin{equation}
\label{eq:modExpanded}
y_i
= \beta_0
+
\sum_{j=1}^{p} x_{ij}\beta_j
+
\sum_{j_1, j_2} x_{ij_1}x_{ij_2}\beta_{j_1 j_2}
+
\cdots+
\sum_{j_1, j_2,\ldots,j_k} \left(\prod_{\ell=1}^k x_{ij_\ell}\right)\beta_{j_1 j_2\cdots j_k}
+
\varepsilon_i\,\,,
\end{equation}
where $\beta_j$ denotes the main effect associated with predictor $x_j$, $\beta_{j_1,\cdots,j_\ell}$ denotes the interaction effect between predictors $x_{j_1},\cdots,x_{j_\ell}$, and $\varepsilon_i$ represents the error term. Higher-power terms arise whenever predictor indices $j_\ell$ are equal. 

To aid in DAG construction, we will express this model using a polynomial basis expansion indexed by multi-indices. Let $\bm x = (x_1, \dots, x_p)$ denote a vector of predictors. Define a multi-index 
\[
\bm\alpha = (\alpha_1, \dots, \alpha_p) \in \mathbb{N}_0^p\,\,,
\]
and the associated monomial basis function
\[
\mathrm{x}^{\bm\alpha} = \prod_{j=1}^{p} x_j^{\alpha_j}\,\,.
\]
To allow different predictors to have different maximum polynomial degrees, let $d_j$ denote the maximum allowed degree for predictor $x_j$. The set of admissible polynomial terms is then indexed by

\[
\mathcal{A} =
\left\{
\bm\alpha \in \mathbb{N}_0^p |
0 \le \alpha_j \le d_j
\right\}\,,\,\,\,\,\,j=1,\,2,...,\,p\,\,.
\]
Using this basis, the regression model can be written as

\begin{equation}
\label{eq:modMonomial}
y_i =
\beta_0 +
\sum_{\bm\alpha \in \mathcal{A} \setminus \{0\}}
\beta_{\alpha} \mathrm{x}_i^{\bm\alpha}
+
\varepsilon_i\,\,,
\end{equation}
where $\beta_\alpha$ denotes the regression coefficient associated with basis function $\mathrm{x}^{\bm\alpha}$.

\subsection{Comparable Approaches}
\label{ss:comparableApproaches}
Regression modeling under heredity constraints, where the inclusion or influence of higher-order terms depends on lower-order parent terms, has been studied extensively. Among shrinkage-based approaches, however, the primary competitors to ILS are the frequentist hierarchical lasso of \cite{bien2013lasso} and the Bayesian hierarchical shrinkage framework of \cite{griffin2017hierarchical}.

Both methods are most naturally formulated in low-order polynomial regression settings, where interactions and quadratic terms induce a relatively simple hierarchical structure. Consequently, we introduce these approaches in the context of a second-order polynomial regression model, which serves as a special case of Equation~\ref{eq:modExpanded}. 

\begin{equation}
y_i
=
\beta_0
+
\sum_{j=1}^{p} x_{ij}\beta_j
+
\sum_{j,k} x_{ij}x_{ik}\beta_{jk}
+
\varepsilon_i\,\,,
\end{equation}
This setting is not merely expository convenience. Rather, it reflects the fact that extending existing heredity-constrained methods to deeper polynomial hierarchies generally requires increasingly elaborate constraints, grouping structures, or prior specifications as the order of the expansion grows. 

\subsubsection{Hierarchical Lasso}

The hierarchical lasso of Bien et al. (2013) is a convex regularization method that enforces strong heredity by coupling each interaction coefficient to its associated main effects \cite{bien2013lasso}. The method estimates coefficients by minimizing a penalized least-squares objective subject to constraints that ensure interaction terms can be nonzero only when both parent effects are present. The optimization problem is 
\begin{equation*}
\min_{\bm{\beta},\theta}
q(\bm\beta,\Theta)
+ \lambda \|\bm{\beta}\|_1
+ \frac{\lambda}{2}\|\Theta\|_1\,\,,
\end{equation*}
where $\bm\beta$ are the main effects, $\Theta\in \mathbb{R}^{p \times p}$ is the matrix of interaction terms such that $\Theta_{jk}=\Theta_{kj}=\beta_{jk}$ , and $\lambda\geq0$ controls the influence of the data vs the $\ell_1$ penalty. To simplify notation, quadratic terms are excluded in the standard formulation, making $\Theta_{jj}=0\,\forall\, j$. However, the method has the ability to model quadratic terms (this was employed in the Section \ref{s:compareSims} simulation study).

The optimization is subject to the hierarchy constraints $\|\theta_j\|_1 \leq |\beta_j|$, where $\theta_j$ denotes the $j$th row of $\Theta$. These constraints ensure that interactions cannot enter the model without the presence of their associated main effects.

To obtain a convex formulation, the coefficients are decomposed into positive and negative parts$\beta_j = \beta_j^+ - \beta_j^-$ with $\beta_j^+, \beta_j^- \ge 0$, leading to the convex problem

\begin{equation*}
\min_{\beta_j^+,\beta_j^-,\Theta}
q(\beta_j^+ - \beta_j^-, \Theta)
+ \lambda \sum_j (\beta_j^+ + \beta_j^-)
+ \frac{\lambda}{2}\|\Theta\|_1\,\,,
\end{equation*}
subject to $\|\theta_j\|_1 \le \beta_j^+ + \beta_j^-$. This formulation preserves convexity, allowing the optimization problem to be solved efficiently even in moderately high-dimensional settings while guaranteeing hierarchical structure with probability one \cite{bien2013lasso}. 

The method produces sparse and interpretable models and serves as a natural frequentist benchmark for hierarchical interaction modeling. However, the framework is primarily designed for pairwise interactions. Extending it to higher-order interactions or deeper hierarchical structures requires introducing additional groups and constraints, substantially increasing implementation complexity.

\subsubsection{Hierarchical Bayesian Shrinkage}

As a foundational example of Bayesian global-local shrinkage, consider the linear regression model with horseshoe prior shrinkage \cite{carvalho2010} 
\begin{equation}
\label{eq:HSpriors}
\begin{aligned}
\bm y \mid \bm \beta,\sigma^2
    &\sim N_n\left(\bm X\bm \beta,\,
    \sigma^2\bm I_n\right) \\
\bm\beta^* \mid \sigma^2,\tau^2,\bm\lambda
    &\sim N_p\left(\bm 0,\,
    \sigma^2\tau^2 D_{\bm\lambda}\right) \\
\tau &\sim C^+(0,1) \\
\lambda_j &\sim g^+(\lambda_j)\,\,, 
\end{aligned}
\end{equation}
where $n$ is the sample size, $\bm X$ is a $n \times p$ matrix of $p$ predictors, $\bm \beta^*$ is the set of regression coefficients excluding the intercept, $\tau^2$ is the global shrinkage parameter, $D_\lambda = \text{diag}\left(\lambda_j^2: j=1,2,\ldots, p\right)$ is a diagonal matrix of the local shrinkage parameters $\lambda^2_j$, $C^+(0,\,1)$ denotes a half-Cauchy distribution, and $g^+(\cdot)$ is some distribution with support $\mathbb{R}^+$. 

The standard horseshoe prior assigns independent half-Cauchy local scales so that $g^+(\cdot)$ corresponds to $ C^+(0,1)$ for $j = 1, \ldots, p$ \cite{carvalho2010}. Extensions such as Horseshoe+ and the regularized horseshoe have proposed alternative specifications for the local shrinkage distribution to modify the prior's behavior near zero and in the tails  \cite{Bhadra2017,piironen2017hyperprior}. The ILS methodology can likewise be viewed as a particular specification of the local scale prior $g^+(\cdot)$ , in which the local shrinkage parameters are constrained to satisfy strong heredity conditions.

Griffin and Brown propose a similar framework for respecting strong heredity in which the prior variance of each coefficient is defined as a function of the prior variances of its parent coefficients \cite{griffin2017hierarchical}. Their prior specification for an order-two polynomial model is
\begin{equation*}
\label{eq:GriffinBrownPriors}
\begin{aligned}
\beta_j,\,\beta_{jj} &\sim N(0, \eta_j\gamma_1d) \\
\beta_{jk,j\neq k} &\sim N(0, \eta_j \eta_k \eta_{jk}\gamma_2d) \\
\eta_j &\sim GG\left(\gamma_1,c,\frac{c-1}{\gamma_1}\right)\\
\eta_{jk} &\sim \frac{c}{\gamma_2}GG\left(\gamma_2,c,\frac{c-1}{\gamma_2}\right)\,\,,
\end{aligned}
\end{equation*}
where $\gamma_ld=\operatorname{Var}\!\left[\beta_j^{(l)}\right]$ are level-specific scales, and $\eta \sim GG(\gamma,c,d)$ denotes a gamma-gamma distribution with density
\newpage
\[
g(\eta)
=
\left(\frac{1}{d}\right)^\gamma
\frac{\Gamma(\gamma + c)}{\Gamma(\gamma)\Gamma(c)}
\eta^{\gamma-1}
\left(1 + \frac{\eta}{d}\right)^{-(\gamma+c)}\,\,.
\]
Constraining $\gamma_2= r\gamma_1$ where $r\in (0,1)$ allows higher-order terms to be shrunk more aggressively than lower-order terms while preserving heredity relationships. 

The method is in principle highly flexible and provides a Bayesian mechanism for propagating shrinkage through a hierarchy. However, prior specification becomes increasingly cumbersome as the structure grows deeper or more irregular, since each new dependency pattern requires custom variance relationships and tuning parameters. In a second-order polynomial basis example given by Griffin and Brown, desirable tail behavior was obtained through the choice of the following hyperpriors \cite{griffin2017hierarchical}:
\begin{equation*}
\label{eq:GriffinBrownHyperpriors}
\begin{aligned}
p(d)&\propto(1+d)^{-2}\\
c&=2\\
\gamma_1&\sim \text{Exp}(1)\\
r&\sim \text{Be}(2,6)\,\,\,.
\end{aligned}
\end{equation*}

An extension to a third-order polynomial basis would require additional prior specifications for: 
(1) higher-order regression coefficients such as $\{\theta_{jkm}, \theta_{jjk}, \theta_{jjj} \mid j,k,m \text{ distinct}\}$,
(2) additional variance components $\eta_{ijk}$, and 
(3) new constraints on $(\gamma_1,\gamma_2,\gamma_3)$ to preserve strong heredity. As the polynomial order increases, the number and complexity of these prior specifications grow rapidly, making hyper-parameter tuning increasingly difficult for deep inheritance structures.

These practical difficulties motivate the methodology presented in the following section. Rather than introducing additional variance components and tuning parameters for each successive level of the hierarchy, we sought a Bayesian hierarchical specification that preserves strong heredity through a single, automatic mechanism. Extending the model to deeper or more irregular inheritance structures should require only specifying the predictor hierarchy, rather than redesigning the prior. The ILS framework introduced next was designed to achieve this goal.

\section{Methods}
\label{ch:soln}
While ILS has applications in a wide variety of regression settings, we begin with the polynomial basis because it provides a familiar and interpretable setting in which hierarchical structure arises naturally. Polynomial expansions are widely used for modeling nonlinear relationships, and their algebraic structure induces a clear parent-child relationship among terms. This makes them a convenient framework for illustrating heredity constraints and the operation of the ILS mechanism.
\newpage
To formalize this hierarchical structure, we define a parent-child relationship among the multi-indices given in Equation \ref{eq:modMonomial}. Let $e_j$ denote the $j$th unit vector in $\mathbb{R}^p$. The set of parents of a term indexed by $\bm\alpha$ is defined as
\[
\mathrm{pa}(\bm\alpha) =
\left\{
\bm\alpha - e_j | \alpha_j > 0
\right\}\,,\,\,\,\,\,j=1,\,2,...,\,p\,.
\]

Each parent corresponds to reducing the exponent of a single predictor by one. For example, $x_1^3$ has parent $x_1^2$, and $x_1^2 x_2$ has parents $x_1 x_2$ and $x_1^2$. Notably, the set of predictors with empty parent sets (i.e. main effects) is 

\[
\left\{\mathrm{x}^{\bm\alpha^*}|\bm\alpha^* \in \mathcal{E}_p\right\}\,,\,\,\,\text{where } \mathcal{E}_p = \{e_1, \ldots, e_p\}.
\]

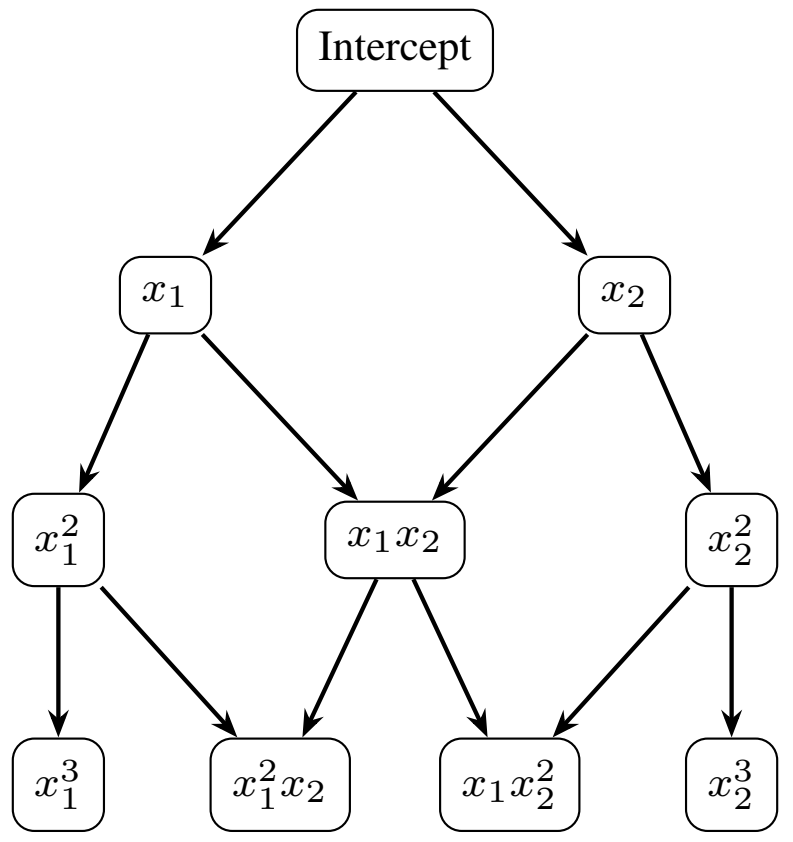
\begin{figure}[H]
\centering
\resizebox{0.32\textwidth}{!}{
\begin{tikzpicture}[
    every node/.style={
        draw,
        rounded corners,
        align=center,
        minimum width=.5cm,
        minimum height=.5cm,
        font=\scriptsize
    },
    arrow/.style={-{Stealth[length=1.8mm]}, thick}
]

\node (b0) at (0,1.6) {Intercept};

\node (x1) at (-1.5,0) {$x_1$};
\node (x2) at (1.5,0) {$x_2$};

\node (x1sq) at (-2.2,-1.6) {$x_1^2$};
\node (x1x2) at (0,-1.6) {$x_1x_2$};
\node (x2sq) at (2.2,-1.6) {$x_2^2$};

\node (x1cu) at (-2.2,-3.2) {$x_1^3$};
\node (x1sqx2) at (-0.75,-3.2) {$x_1^2x_2$};
\node (x1x2sq) at (0.75,-3.2) {$x_1x_2^2$};
\node (x2cu) at (2.2,-3.2) {$x_2^3$};

\draw[arrow] (b0) -- (x1);
\draw[arrow] (b0) -- (x2);

\draw[arrow] (x1) -- (x1sq);
\draw[arrow] (x1sq) -- (x1cu);

\draw[arrow] (x2) -- (x2sq);
\draw[arrow] (x2sq) -- (x2cu);

\draw[arrow] (x1) -- (x1x2);
\draw[arrow] (x2) -- (x1x2);

\draw[arrow] (x1sq) -- (x1sqx2);
\draw[arrow] (x1x2) -- (x1sqx2);

\draw[arrow] (x2sq) -- (x1x2sq);
\draw[arrow] (x1x2) -- (x1x2sq);

\end{tikzpicture}
}
\caption{Inheritance DAG for a degree-3 polynomial surface with two main effects.}
\label{fig:polybasis}
\end{figure}

These relationships induce a directed acyclic graph (DAG) whose nodes correspond to polynomial basis functions and whose edges are defined by $(\bm\alpha - e_j) \rightarrow \bm\alpha$. We call this an \textit{inheritance DAG}. Thus, higher-order polynomial terms are children of their corresponding lower-order terms. An example inheritance DAG for the polynomial basis is given in Figure \ref{fig:polybasis}.

In the context of this polynomial basis, the strong heredity constraint can be written

\[
\beta_{\bm\alpha} \neq 0
\;\Rightarrow\;
\beta_{\bm\alpha'} \neq 0
\quad
\forall \bm\alpha' \in \mathrm{pa}(\bm\alpha).
\]

\noindent Equivalently, the support of the coefficient vector must be downward closed in the heredity DAG. A model satisfying these conditions is referred to as \emph{well-formulated} \cite{Peixoto1990}.

The goal of ILS is to apply shrinkage within a well-formulated model while preserving strong heredity conditions. Consequently, a model that is well-formulated remains well-formulated under shrinkage. This is accomplished by applying an ordering constraint on the local shrinkage parameters corresponding to the model's inheritance structure.

Let $\mathcal{A}$ denote the index set of polynomial basis functions in a well-formulated model and let $\mathrm{pa}(\bm\alpha)$ denote the parent set which induces an inheritance DAG. For each coefficient $\beta_{\bm\alpha}$, let $\lambda_{\bm\alpha}^2$ denote the associated local shrinkage parameter in the global-local prior. Smaller values of $\lambda_{\bm\alpha}^2$ correspond to stronger shrinkage of $\beta_{\bm\alpha}$ toward zero.

\begin{definition}[Inherited Local Shrinkage]
A collection of local shrinkage parameters $\{\lambda_{\bm\alpha}^2 : \bm\alpha \in \mathcal{A}\}$ satisfies the \emph{inherited local shrinkage} (ILS) condition if, for every $\bm\alpha \in \mathcal{A}$ with at least one parent,
\[
\lambda_{\bm\alpha}^2 \le
\min_{\bm\alpha' \in \mathrm{pa}(\bm\alpha)} \lambda_{\bm\alpha'}^2.
\]
\end{definition}

ILS ensures that the local shrinkage parameter of any child coefficient cannot exceed that of its parents. The following property formalizes the relationship between ILS and the heredity structure of the model.

\begin{proposition}[Shrinkage propagation]
Suppose the local shrinkage parameters satisfy the inherited local shrinkage condition. If a parent coefficient $\beta_{\bm\alpha'}$ is strongly shrunk toward zero through a small local shrinkage parameter $\lambda_{\bm\alpha'}^2$, then every descendant coefficient $\beta_{\bm\alpha}$ satisfying $\bm\alpha' \in \mathrm{pa}^*(\bm\alpha)$ must receive at least as much shrinkage, where $\mathrm{pa}^*(\bm\alpha)$ denotes the set of all ancestors of $\bm\alpha$ in the heredity DAG.
\end{proposition}

More formally, it can be shown that the absolute value of a descendant coefficient is stochastically dominated by those of its ancestors (see Proposition \ref{prop:ILSdominance} and its proof in Appendix A).

Consequently, shrinkage applied to lower-order polynomial terms automatically propagates forward through the DAG to any higher-order terms that depend on them. This property ensures that models remain well-formulated under shrinkage, as desired. The contrapositive is also notable: if a higher-order term is high-information and thus not shrunk, its entire parent set is preserved from shrinkage, producing a back-propagation of high prior variance through the DAG. 

If we define the upper bounds
\[
m_{\bm\alpha}
=
\begin{cases}
\min_{\bm\alpha' \in \mathrm{pa}(\bm\alpha)} \lambda_{\bm\alpha'}^2\,\,, & \bm\alpha\notin\mathcal{E}_p, \\
\infty\,\,, &  \bm\alpha\in\mathcal{E}_p \,\,\,,
\end{cases}
\]

\noindent then the ILS priors are given by Equation \ref{eq:HSpriors} with $g^+\left(\lambda_{\bm\alpha}\right)$ corresponding proportionally to $\mathrm{C}^+(0, 1)\mathbbm{1}_{\left\{\lambda_{\bm\alpha} \in (0,m_{\bm\alpha})\right\}}$. The conditional posteriors for $\bm{\beta}$, $\sigma^2$, and $\tau$ align with the standard horseshoe prior result given in \cite{Makalic2016}. 

Markov chain Monte Carlo (MCMC) sampling of $[\bm\lambda|\,\cdot\,]$ can be done in a single block in an ordering that traverses the inheritance DAG. The acyclicity of DAGs ensures that there exists a total ordering of the $q$ nodes $\prec$ such that $\bm\alpha^{(1)}\prec\bm\alpha^{(2)}\prec \cdots\prec\bm\alpha^{(q)}$. Thus, the local shrinkage joint posterior can be written
\begin{equation}
\label{eq:jointLocalShrk}
[\bm\lambda|\,\cdot\,]= [\lambda_{\bm\alpha^{(1)}}|\,\cdot\,][\lambda_{\bm\alpha^{(2)}}|\lambda_{\bm\alpha^{(1)}},\,\cdot\,]\cdots[\lambda_{\bm\alpha^{(q)}}|\lambda_{\bm\alpha^{(1)}},\cdots,\,\lambda_{\bm\alpha^{(q-1)}},\,\cdot\,]\,\,\,.
\end{equation}
The local shrinkage for main effects $[\lambda_{\bm\alpha} | m_{\bm\alpha}=\infty,\,\cdot\,]$ are unconstrained and thus follow the standard horseshoe prior form \cite{Makalic2016}. In the case where the parent set is nonempty,
\begin{equation*}
\begin{aligned}
\lambda_{\bm\alpha}^2
\mid m_{\bm\alpha}<\infty,\,
\cdot\;
&\sim\;
\operatorname{IG}
\left(
1,\,
\frac{\beta_{\bm\alpha}^2}
{2\sigma^2\tau^2}
+
\frac{1}{\nu_{\bm\alpha}}
\right)
\mathbbm{1}_{(0,m_{\bm\alpha})}(\lambda_{\bm\alpha}^2)\\
\nu_{\bm\alpha}|\lambda_{\bm\alpha}^2\;
&\sim\;
\operatorname{IG}\left(
1,\,1+
\frac{1}
{\lambda_{\bm\alpha}^2}
\right)\;,
\end{aligned}
\end{equation*}
where $\operatorname{IG}(\cdot)$ corresponds to the inverse-gamma distribution.

\subsection{Implementation} \label{ss:Implimentation}
 
In addition to the standard regression inputs, ILS requires the user to specify the inheritance DAG. In practice, this structure is most conveniently provided as a list identifying the direct parents of each term in the design matrix. If Markov chain Monte Carlo sampling of local shrinkage is to be carried out as shown in Equation \ref{eq:jointLocalShrk}, the elements of the parent list can be ordered according to the total ordering $\prec$.

The design matrix should be well-formulated by construction with standardized predictors. The model is typically specified with more terms than are expected to be necessary, allowing the shrinkage prior to adaptively suppress unsupported coefficients while preserving strong heredity. 

For categorical predictors represented by multiple design matrix columns, all associated coefficients should share a common local shrinkage parameter so that the factor is regularized as a single unit. This reduces the number of local shrinkage parameters below the number of design matrix columns, necessitating an indexing vector that maps each column to its corresponding shrinkage parameter. Although motivated here by categorical predictors, the same framework can be used whenever scientific or structural considerations suggest that a collection of predictors should be regularized as a group.

\section{Simulation Studies}
\label{ch:sim}

We conducted two simulation studies using the polynomial basis. The first study compared ILS with two alternative models that incorporate strong inheritance, as well as a horseshoe prior model without an inheritance structure. The second study demonstrated ILS's ability to recover 
functional form in a more complex model space.

\subsection{Data Generation}
The two simulations employed identical data generation procedures. The main difference was that the first simulation used 500 datasets for Monte Carlo estimation, while the second simulation utilized a single dataset for a demonstration of ILS's performance in a deeper inheritance structure. The response was generated so that the regression coefficients were calibrated to achieve a prespecified signal-to-noise ratio (SNR) while distributing signal equally across all active terms. The simulation code constructed the coefficient vector and error variance using only the predictors designated as truly nonzero, which we call \textit{active predictors}. 

Let $X_S$ denote the design matrix containing only the active predictors, and let $\tilde X_S = [\mathbf{1}, X_S]$ include an intercept column. Suppose there are $q$ active predictors. The response was generated according to
\[
y = \tilde X_S \bm{\beta} + \varepsilon,
\qquad
\varepsilon \sim N(0,\sigma^2 I_n),
\]
where both the coefficient vector $\bm{\beta}$ and the error variance $\sigma^2$ were chosen to satisfy two conditions:
\begin{enumerate}
    \item The intercept coefficient is fixed at 1.
    \item The SNR is set to the user-specified value.
\end{enumerate}

The active predictors may be correlated, so assigning identical weights directly to the coefficients would not produce equal contributions in the fitted linear predictor. To account for the correlation structure, the simulation used the QR decomposition of the active design matrix, $\tilde X_S = QR$, where $R$ is upper triangular. The coefficient vector was then defined as $\bm{\beta} = cR^{-1}\bm\eta$, where $c$ is a scalar obtained via optimization that ensured the intercept coefficient $\beta_0 = 1$ and $\bm\eta$ controlled how information is distributed among the active predictors. This transformation maps equal information to a coefficient vector in a way that respects the geometry of the design matrix. Thus, the effective signal was spread evenly over the active predictors after accounting for collinearity.

Once $\bm{\beta}$ was obtained, the variance of the linear predictor was estimated using the sample covariance matrix of the active design matrix, denoted $\Sigma_X$. The signal variance was computed as $\mathrm{Var}(\tilde X_S \bm{\beta})=\bm{\beta}^\top \Sigma_X \bm{\beta}$. To achieve the desired signal-to-noise ratio, $\mathrm{SNR}=\mathrm{Var}(\tilde X_S \bm{\beta})\sigma^{-2}$,
the error variance was set to
\[
\sigma^2
=
\frac{\bm{\beta}^\top \Sigma_X \bm{\beta}}{\mathrm{SNR}}.
\]
Larger SNR values therefore correspond to smaller residual variance and an easier variable-selection problem. Finally, the response was simulated as
\[
y_i
=
\beta_0
+
\sum_{j \in S} x_{ij}\beta_j
+
\varepsilon_i,
\qquad
\varepsilon_i \stackrel{\mathrm{iid}}{\sim} N(0,\sigma^2),
\]
with all inactive predictors assigned coefficients equal to zero.

\subsection{Simulation Experiment 1: Comparison to Existing Methods}\label{s:compareSims}
\begin{figure}[H]
    \centering
    \includegraphics[width=\linewidth]{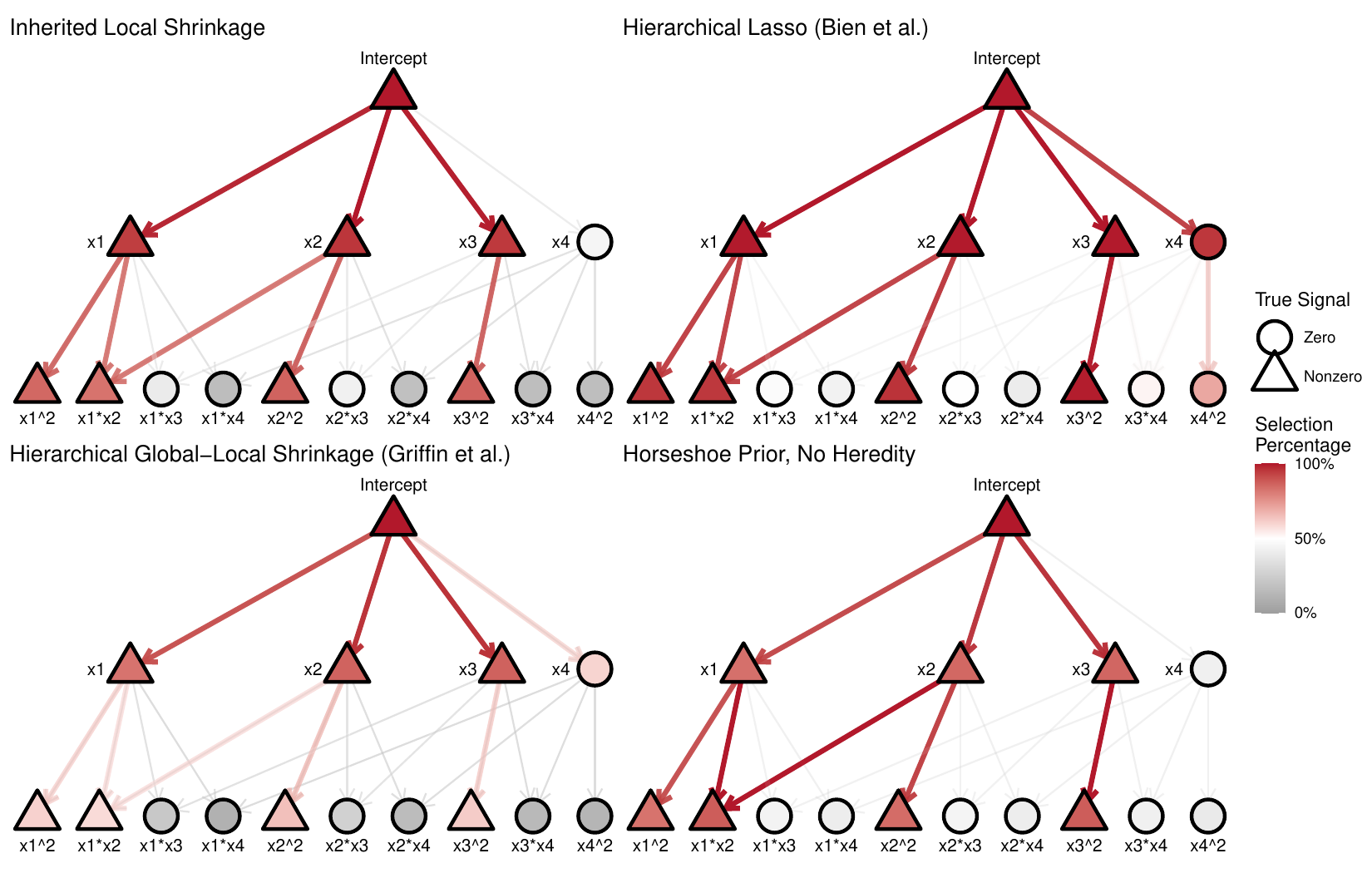}
    \caption{DAG summaries showing the proportion of 500 Monte Carlo simulations in which each predictor was selected by each method. Red graph edges indicate variables selected in more than 50\% of simulations, with darker shades corresponding to higher selection frequencies. Node shapes indicate the true status of each term in the data-generating process.}
    \label{fig:DAG_MC_Results}
\end{figure}

The simulation study was carried out in an order-two polynomial model space in which a well-formulated subset of terms were chosen as active predictors. Four methods were considered: ILS, the hierarchical lasso from \cite{bien2013lasso} (HL), the hierarchical global-local shrinkage model from \cite{griffin2017hierarchical} (HGLS), and a horseshoe prior model with no heredity (HS). The first three models respect strong heredity, while the fourth demonstrates a straightforward application of shrinkage where heredity is disregarded. Each method was applied to 500 simulated data sets of size $n=30$ with $\mathrm{SNR}=4$ and $\bm\eta=\bm1$, allocating equal information to all significant variables. The set of significant predictors was held constant over all simulations, but each simulated data set had a unique design matrix and response vector. The model space and true model are shown in Figure \ref{fig:DAG_MC_Results}. 

A comparison methodology had to be devised that would make the methods comparable, taking into account the frequentist nature of HL and the Bayesian nature of ILS, HGLS, and HS. We chose to use the fitted functional form of each model as the standard for comparison. For HL, this was the set of predictors with non-zero regression coefficient estimates. For the Bayesian methods, we used the principle of the \textit{half-thresholding estimator} from \cite{tang2018bayesian} to define the set of predictors that have been ``selected" by each of the three shrinkage methods for each simulated data set. A predictor was considered selected if
$$ \left |\hat{\beta}_i^{PM}/\hat{\beta}_i^{OLS}\right |>0.5 \,\,\,,$$
where $\hat{\beta}_i^{PM}$ is the posterior mean of $\beta_i$ under a shrinkage prior and $\hat{\beta}_i^{OLS}$ is the ordinary least squares estimate. Having reduced both the frequentist lasso and Bayesian shrinkage models to discrete selection output, we recorded the set of selected predictors (i.e. the recovered functional form) at each iteration for each model. Thus, Monte-Carlo estimates of true positive rates, false positive rates, false detection rates, accuracy, and Jaccard indices were obtained by computing means over the 500 simulated data sets. These results are given in Table \ref{tab:MCResults}.

\begin{table}[h]
    \centering
    \small
    \begin{tabular}{lccccc}
        \hline
        Model & TPR & FPR & FDR & ACC & Jaccard \\
        \hline
        Inherited Local Shrinkage
            & \textbf{0.894}
            & \textbf{0.280}
            & \textbf{\textcolor{red}{0.239}}
            & \textbf{\textcolor{red}{0.807}}
            & \textbf{\textcolor{red}{0.698}} \\
        Hierarchical Lasso (Bien et al.)
            & \textbf{\textcolor{red}{0.979}}
            & 0.571
            & 0.368
            & 0.704
            & \textbf{0.623} \\
        Hierarchical Global-Local Shrinkage (Griffin et al.)
            & 0.716
            & \textbf{\textcolor{red}{0.231}}
            & \textbf{0.244}
            & \textbf{0.742}
            & 0.581 \\
        Horseshoe Prior, No Heredity
            & 0.852
            & 0.421
            & 0.331
            & 0.715
            & 0.599 \\
        \hline
    \end{tabular}
    \caption{Mean selection metrics comparing shrinkage methods over 500 Monte-Carlo simulations. TPR: true positive rate, FPR: false positive rate, FDR: false discovery rate, ACC: accuracy, and Jaccard similarity coefficient. Red bold values indicate the best-performing model for each metric, while bold values indicate the second-best performer.}
    \label{tab:MCResults}
\end{table}
\newpage
The high true positive rate shows that HL excels in controlling Type II error, but it had the worst Type I error, as reflected in its high false positive rate. Conversely, HGLS had the best false positive rate, but the worst true positive rate. ILS, on the other hand, was second-best in both metrics, indicating more balanced inference. Furthermore, ILS had the best false detection rate (another Type I error control metric), accuracy, and Jaccard index. These results are illustrated in Figure \ref{fig:DAG_MC_Results}, which shows the proportion of the MC simulations that each predictor was selected in each model.

While HL and HGLS may be good options depending on the data and the researcher's prior knowledge of the model space, this simple polynomial space simulation shows that ILS performs at least comparably to these competitors and may achieve better overall performance through the balance of Type I and Type II error.

\subsection{Simulation Experiment 2: Functional Form Discovery and Shrinkage Propagation}

The second simulation study used four main effects in an order-four polynomial space. The simulation parameters were $n=150$ with $\mathrm{SNR}=4$ and $\bm\eta=\bm1$. Ten thousand MCMC samples were used for estimation after removing fifty thousand as a burn-in. The true model consisted of three lineages of active predictors, chosen in such a way to demonstrate ILS working in various inheritance structures. The lineages were
\begin{enumerate}[label=\Alph*:]
    \item $\left\{x_1,\, x_1^2,\,x_1^3,\, x_1^4\right\}$
    A ``deep" lineage, where a line of direct descendants should all remain unshrunk
    \item $\left\{x_2,\, x_3,\,x_2x_3,\,x_2x_3^2\right\}$
    A ``wide" lineage with some significant interaction terms, but with many related interaction terms that should be shrunk
    \item $\left\{x_4\right\}$
    A ``shallow" lineage, where a main effect should be included, but all descendants shrunk to zero
\end{enumerate}

\begin{figure}[H]
    \centering
\begin{minipage}[c]{0.12\linewidth}
    \centering
    \textbf{Lineage A}
\end{minipage}
\hspace{6em}
\begin{minipage}[c]{0.065\linewidth}
    \centering
    \includegraphics[width=\linewidth]{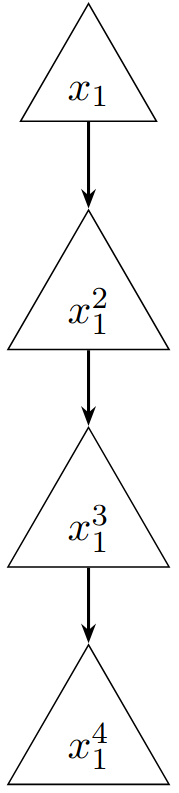}
\end{minipage}\hfill
\begin{minipage}[c]{0.375\linewidth}
    \centering
    \includegraphics[width=\linewidth]{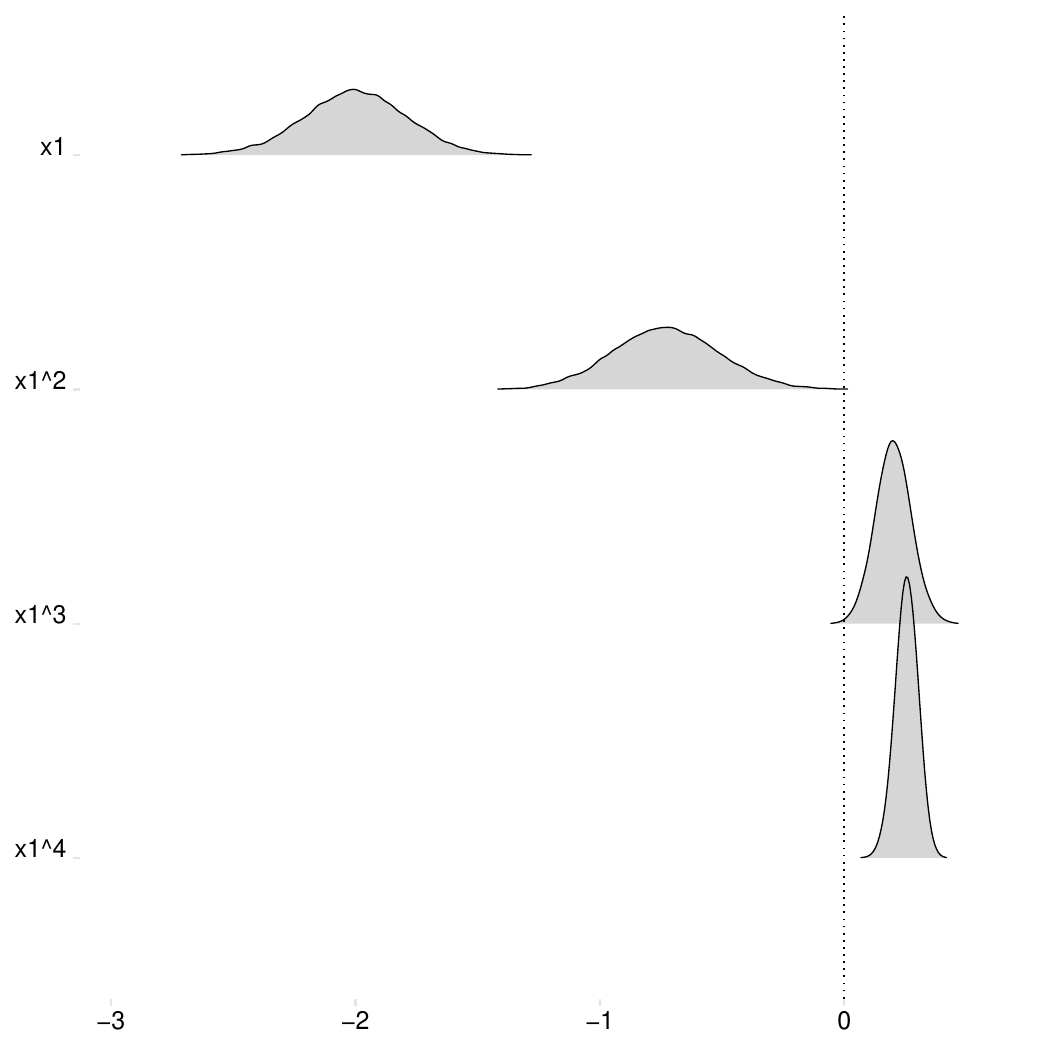}
\end{minipage}

\vspace{0.5em}

\begin{minipage}[c]{0.1\linewidth}
    \centering
    \textbf{Lineage B}
\end{minipage}
\begin{minipage}[c]{0.4\linewidth}
    \centering
    \includegraphics[width=\linewidth]{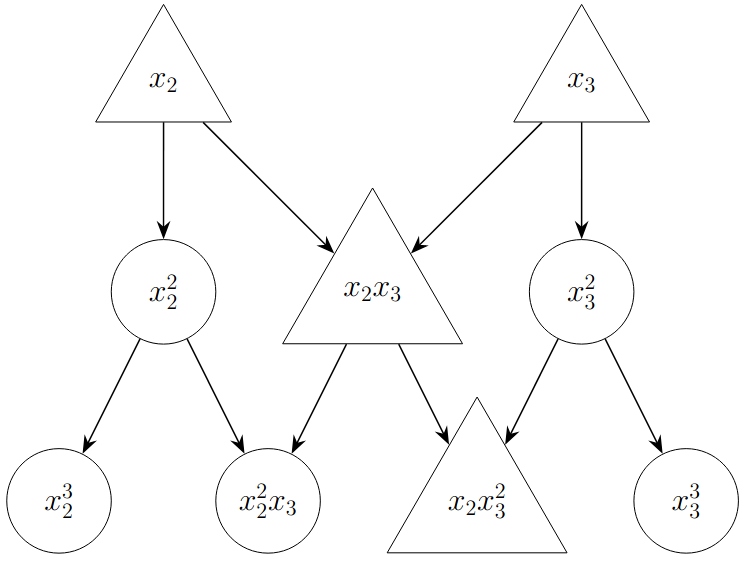}
\end{minipage}\hfill
\begin{minipage}[c]{0.4\linewidth}
    \centering
    \includegraphics[width=\linewidth]{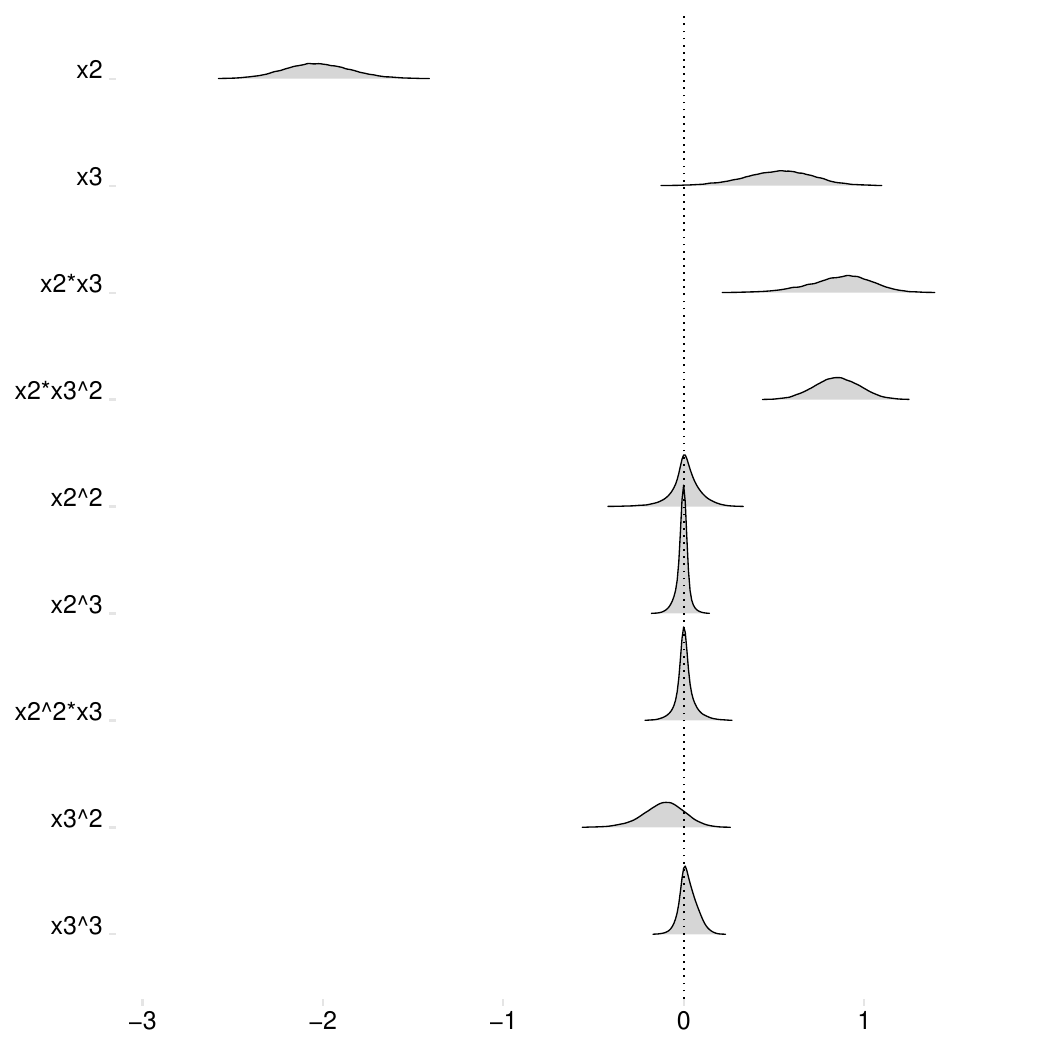}
\end{minipage}

\vspace{0.5em}

\begin{minipage}[c]{0.1\linewidth}
    \centering
    \textbf{Lineage C}
\end{minipage}
\begin{minipage}[c]{0.4\linewidth}
    \centering
    \includegraphics[width=\linewidth]{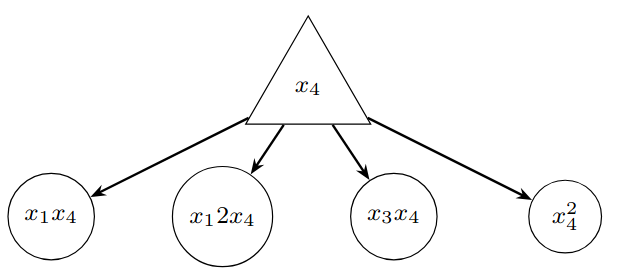}
\end{minipage}\hfill
\begin{minipage}[c]{0.4\linewidth}
    \centering
    \includegraphics[width=\linewidth]{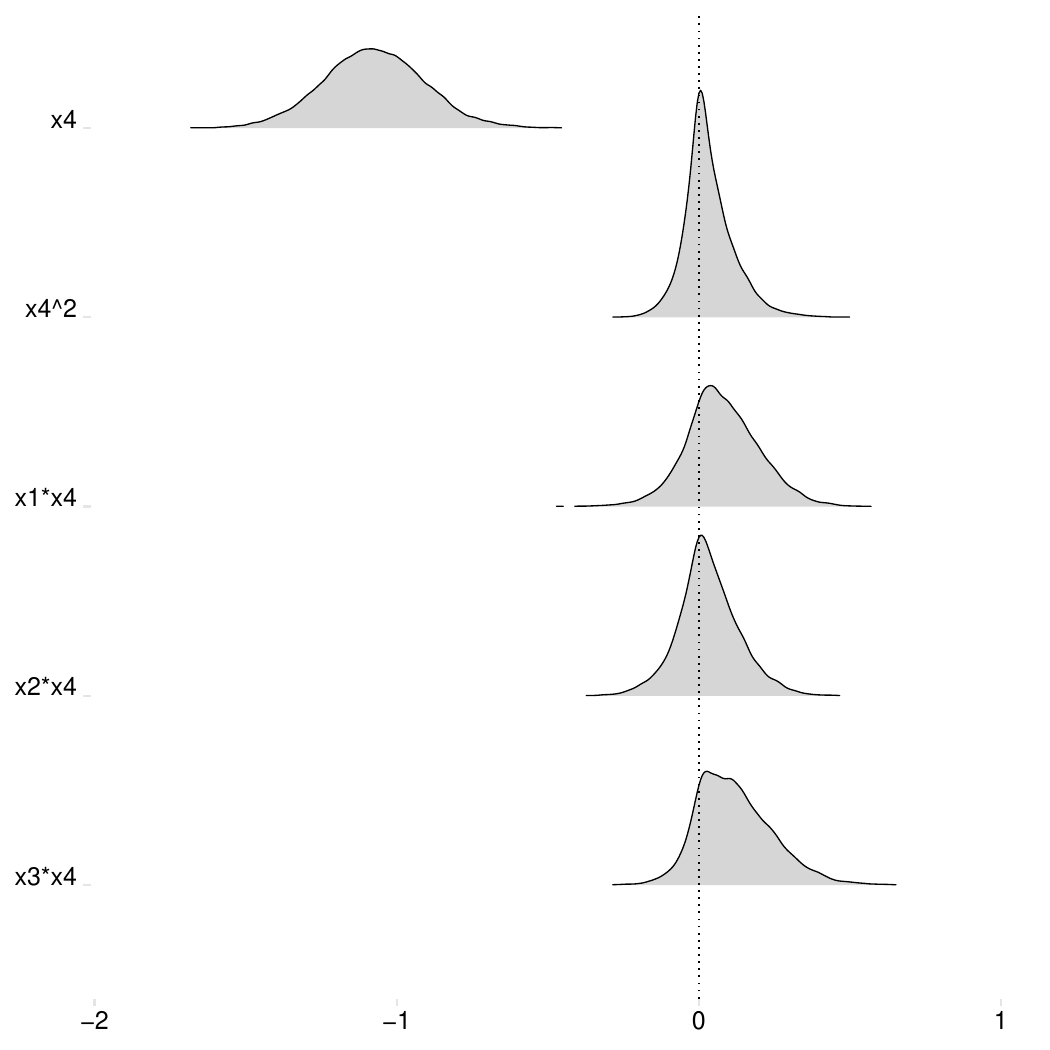}
\end{minipage}
    \caption{Relevant inheritance sub-DAGs for each of the three lineages and their associated estimated regression coefficient posteriors from an ILS model. Nonzero and zero information variables are indicated by triangles and circles respectively.  \textbf{Top:} Lineage A \textbf{Middle:} Lineage B \textbf{Bottom:} Lineage C }
    \label{fig:demonstrationRidge}
\end{figure}

The resulting model contains many predictors, but we focus on examining the sub-DAGs associated with each of the three lineages (Figure \ref{fig:demonstrationRidge}) in order to demonstrate functional form recovery and propagation of shrinkage through the DAG in ILS models.

The ILS model successfully recovered the true functional structure in all three lineages. In Lineage A, the higher-order terms were estimated as clearly nonzero despite the structural tendency of the prior to impose stronger shrinkage on descendants. No spike at zero was observed in their posterior distributions, indicating that sufficiently strong signal overcame the additional shrinkage induced by the hierarchical prior.

Lineage B illustrates how shrinkage is applied selectively to predictors that are both uninformative and disconnected from any active lineage. In this example, the parent term $x_3^2$ was estimated to have a small effect, yet its posterior retained a diffuse, approximately Gaussian shape centered near zero rather than the sharp spike associated with aggressive shrinkage. This behavior reflects the fact that an inactive parent is not automatically shrunk simply because its coefficient is close to zero. Instead, because $x_3^2$ is required to preserve strong heredity for its descendants, its local shrinkage parameter remains relatively large. However, its zero-information descendant $x_3^3$ exhibited concentrated spike-shaped posteriors near zero.

\newpage
Lineage C demonstrates that the converse does not hold: a significant parent does not guarantee protection from shrinkage for its descendants. Although $x_4$ was estimated to be strongly nonzero, all four of its inactive children were heavily shrunk, producing posterior distributions sharply concentrated around zero. Slight posterior skewness in the interaction terms indicate that sampling variability induced a weak nonzero tendency, but this effect was dominated by the shrinkage prior, which pulled the distributions back toward zero.

Together, these results illustrate the defining behavior of ILS: shrinkage propagates forward through the hierarchy, ensuring that inactive branches are pruned while preserving lower-order terms necessary to maintain strong heredity.

\section{Adaptive Multiresolution Modeling with ILS} 
\label{ch:adaptiveRes}

We now consider problems where a fine-resolution inference might be appropriate, but the required resolution is unknown and may vary across the domain of study. While the generality of this problem should not be overlooked, two prevalent applications in this research are analyses of temporal and spatial data.

ILS is particularly well suited for this problem. First, a basis must be chosen in which every child predictor can be considered ``higher resolution" than their parent(s). The researcher then generates predictors up to the highest resolution that they are willing to consider. In practice, the maximum resolution may be determined using prior knowledge and/or computational constraints. When the ILS model is fit, the informative high resolution predictors and all of their parents will not be shrunk while any unnecessarily high resolution bases will be shrunk to zero. 

\subsection{One Dimension: Modeling Unemployment Time Series Using Wavelets}
We first consider one-dimensional modeling using a wavelet basis. As a challenging test case, we analyze a long-term time series of U.S. unemployment rates spanning several decades. This example highlights the utility of ILS when combined with a flexible multi-resolution basis representation.

The wavelet basis approach divides the domain of interest into subregions at each resolution, and models each subregion using a special restricted-domain function called a wavelet \cite{Daubechies1992, Mallat1999}. Using wavelet predictors allows a regression model to account for both global structure and local features such as discontinuities or spatial heterogeneity, which is especially useful when data is generated from a nonstationary temporal or spatial process\cite{DonohoJohnstone1994,Nason2008}. 

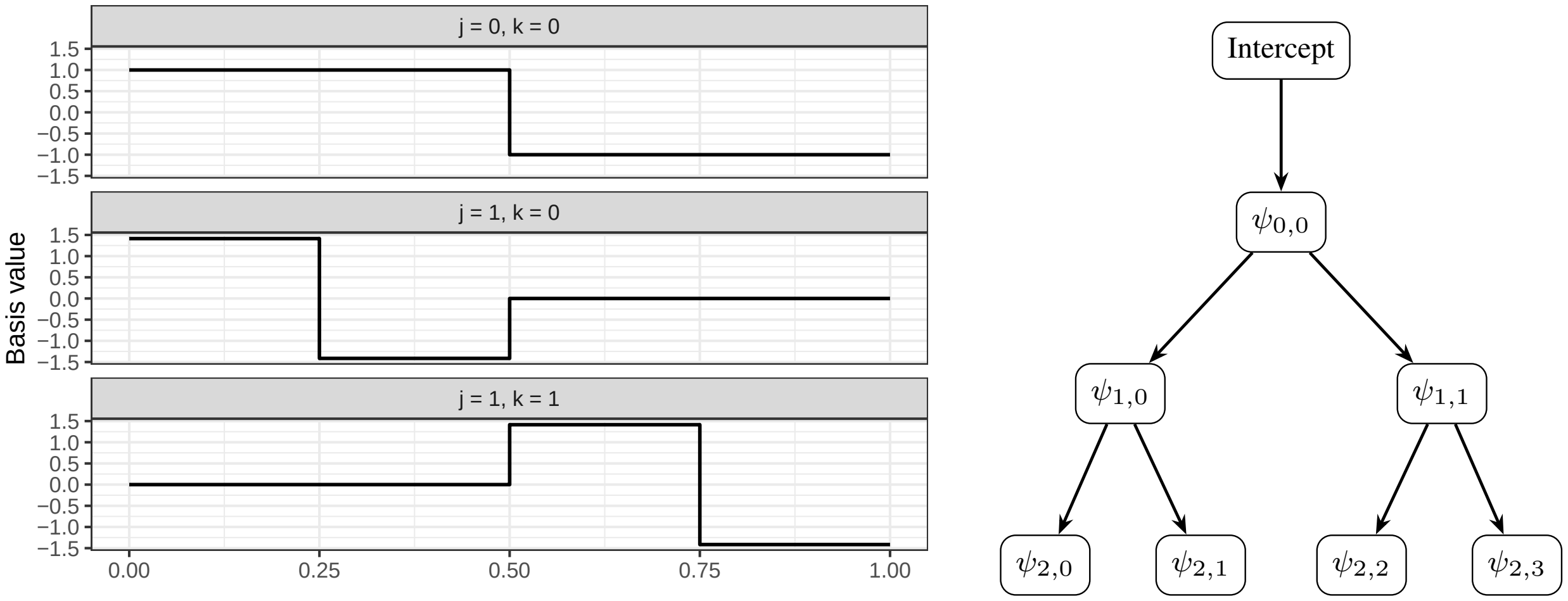
\begin{figure}[H]
    \centering
    \includegraphics[width=.6\linewidth]{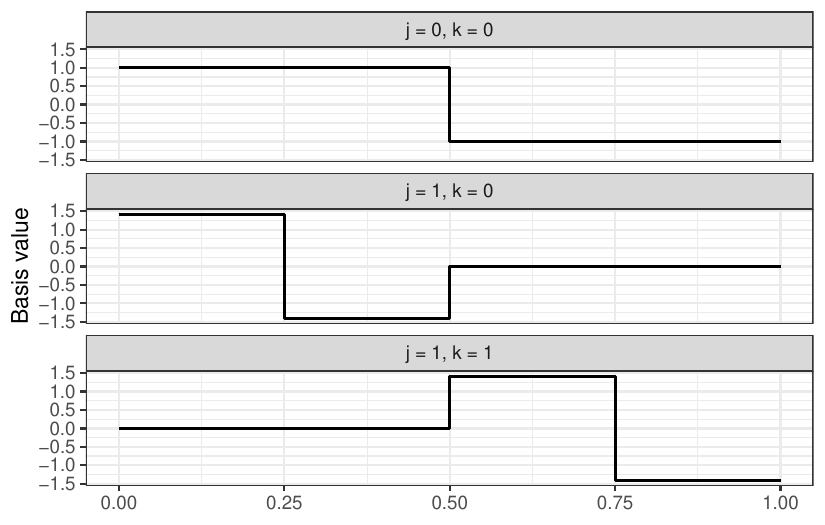}\hfill
    \resizebox{0.37\textwidth}{!}{
\begin{tikzpicture}[
    every node/.style={
        draw,
        rounded corners,
        align=center,
        minimum width=.5cm,
        minimum height=.5cm,
        font=\scriptsize
    },
    arrow/.style={-{Stealth[length=1.8mm]}, thick}
]

\node (b0) at (0,1.6) {Intercept};

\node (r0) at (0,0) {$\psi_{0,0}$};

\node (r10) at (-1.5,-1.6) {$\psi_{1,0}$};
\node (r11) at (1.5,-1.6) {$\psi_{1,1}$};

\node (r20) at (-2.2,-3.2) {$\psi_{2,0}$};
\node (r21) at (-0.75,-3.2) {$\psi_{2,1}$};
\node (r22) at (0.75,-3.2) {$\psi_{2,2}$};
\node (r23) at (2.2,-3.2) {$\psi_{2,3}$};

\draw[arrow] (b0) -- (r0);

\draw[arrow] (r0) -- (r10);
\draw[arrow] (r0) -- (r11);

\draw[arrow] (r10) -- (r20);
\draw[arrow] (r10) -- (r21);

\draw[arrow] (r11) -- (r22);
\draw[arrow] (r11) -- (r23);

\end{tikzpicture}
}
    \caption{\textbf{Left:} Resolutions 0 and 1 of Haar wavelet basis in one dimension over [0, 1]. \textbf{Right:} Inheritance DAG for a one dimensional wavelet basis with dyadic scaling.}
    \label{fig:haar}
\end{figure}

The Haar wavelet is the simplest orthonormal wavelet basis and is particularly attractive for recursive temporal and spatial partitioning \cite{Daubechies1992}. In the context of this work, the Haar wavelet basis produces a model that is both effective at capturing multi-scale trends and illustrative of ILS through its intuitive inheritance structure. The Haar wavelet function is
$$
\psi(t) =
\begin{cases}
1, & 0 \le t < \tfrac{1}{2}, \\
-1, & \tfrac{1}{2} \le t < 1, \\
0, & \text{otherwise}.
\end{cases}
$$
The basis is then generated by $\psi_{j,k}(t) = 2^{j/2} \psi(2^{j} t - k)$, for resolution level $j \in \mathbb{Z}$ and location index $k \in \mathbb{Z}$. The piecewise formulation gives rise to stair-step shapes seen in figure \ref{fig:haar}.

The wavelet basis follows the principle of dyadic scaling, so at each resolution $j>0$, each region in a time series is split into two equally-sized sub-regions (i.e., children regions), and each are modeled using a wavelet basis. 

The recursive partitioning of the study domain induces a natural hierarchical structure: in one dimension, each region at resolution $j$ has two children at the finer-scale resolution $j+1$. When ILS is applied, shrinkage operates along the inheritance graph in a manner that respects the parent-child relationships among coefficients. Specifically, once a wavelet coefficient at a given resolution is deemed uninformative and is strongly shrunk toward zero, all of its descendants at finer resolutions are likewise shrunk toward zero. As a result, the model adaptively determines the finest resolution required in different parts of the study region.

The one-dimensional adaptive resolution example corresponds to the monthly U.S. macroeconomic indicators obtained from the Federal Reserve Economic Data service maintained by the Federal Reserve Bank of St.\ Louis \cite{FRED}. In this analysis, the response variable is \texttt{unemploy}, which records the number of unemployed individuals in the United States each month.

The series spans 1967-2015 and exhibits both smooth long-term trends and abrupt changes associated with business cycles and economic shocks. These characteristics make the series well-suited for evaluating multi-resolution methods, as broad trends can be represented with coarse-scale basis functions while localized fluctuations and structural changes require finer-scale components. For the present analysis, the time index was rescaled to the unit interval and the unemployment series was treated as a one-dimensional signal.

Models were fit using maximum resolution levels ranging from 1 to 9, yielding progressively finer-scale prediction curves. Maximum resolution values greater than 9 were computationally expensive, given the exponential growth of the model space with spatial resolution. Very few high-resolution predictors remained unshrunk at the 9th resolution, which gives some assurance that the global model was not underspecified. Due to the prevalence of skew in shrinkage models, regression coefficients were summarized by their posterior medians. The Markov chain Monte Carlo (MCMC) sample consisted of 1,000 post-burn-in samples. The burn in was the removal of the first 5,000 iterations to ensure convergence. 

\begin{figure}[H] 

    \centering

    \textbf{A}\par
    \includegraphics[width=\linewidth]{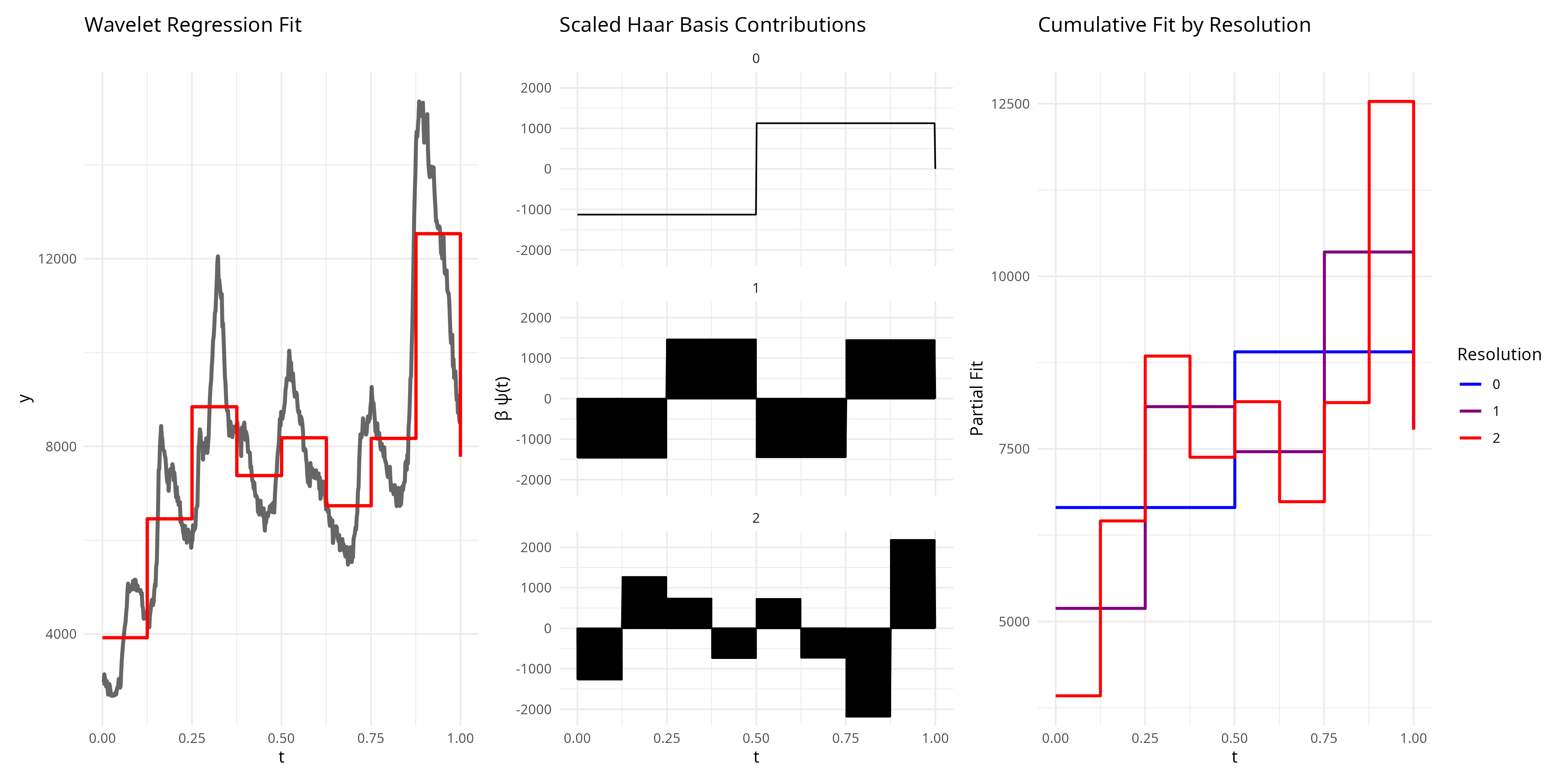}

    \vspace{0.5cm}

    \textbf{B}\par
    \includegraphics[width=\linewidth]{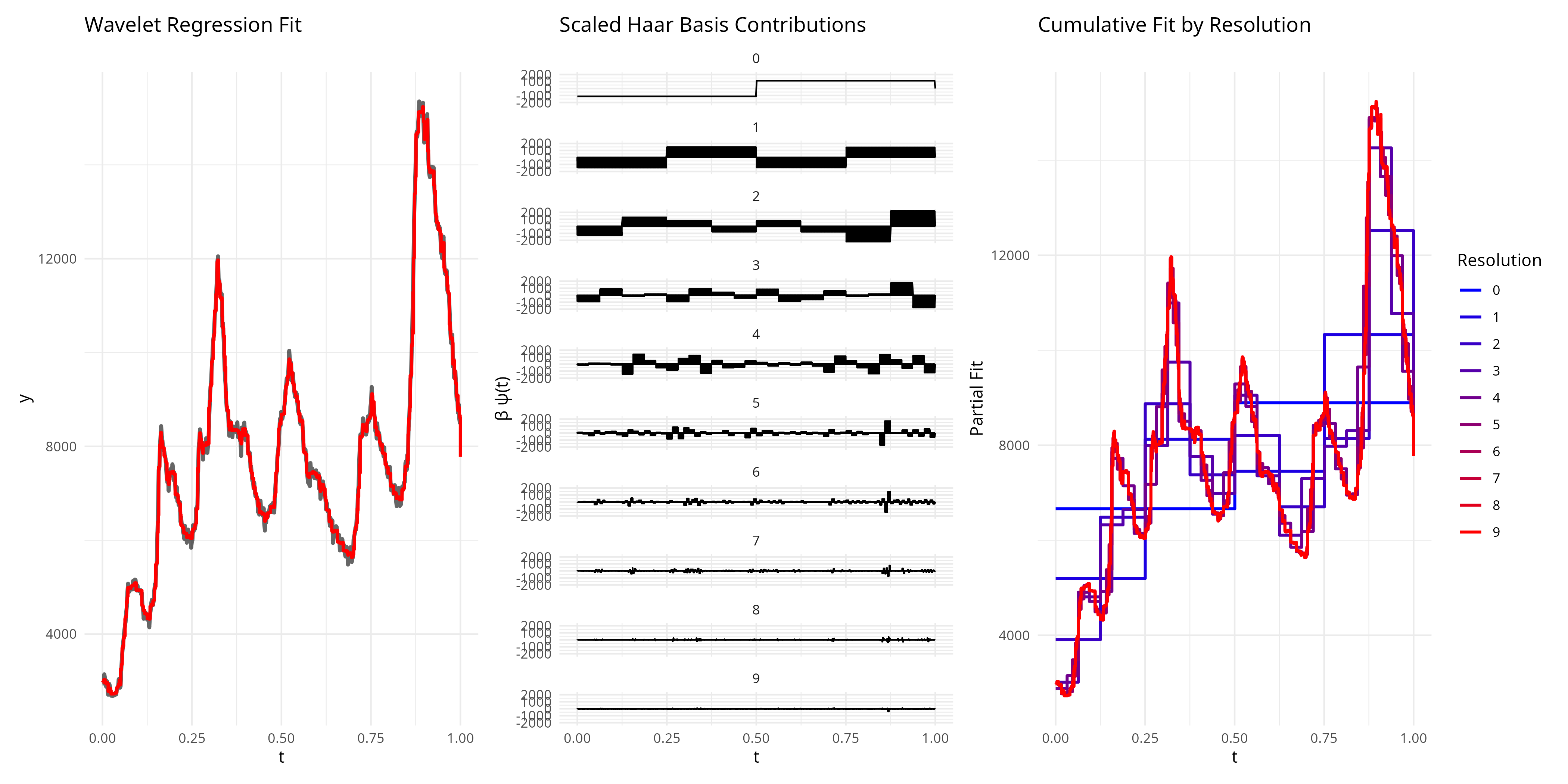}

    \caption{Predictions and scaled Haar basis predictors. Panel A shows an underspecified $J=2$ model and Panel B shows a properly specified $J=9$ model.}
\label{fig:unemploy_J2_J9}
\end{figure}

Figure \ref{fig:unemploy_J2_J9}A illustrates an underspecified model in which all predictors remain largely unshrunk, resulting in a jagged prediction curve that provides a coarse approximation to the underlying signal. This low-resolution fit fails to capture the most extreme response values, leading to the largest prediction errors when unemployment deviates substantially from the sample mean.

The higher resolution model shown in Figure \ref{fig:unemploy_J2_J9}B is much better specified. However, a substantial proportion of subregions did not exhibit strong shrinkage until resolution 8, indicating the prominence of small-scale variation through time. In contrast, most predictors at resolution 9 were heavily shrunk toward zero. The region requiring the finest-scale representation was located near $t = 0.85$. Several intervals required relatively little local refinement, including the interval $(0,\,0.5)$.

\subsection{Two Dimensions: Harvard Forest Digital Terrain Model Data}
\label{ss:2Dsim}
We now demonstrate the use of ILS spatial modeling with two dimensional ecological data. The raster \texttt{HARV\_dsmCrop.tif} is a cropped Digital Surface Model (DSM) for the Harvard Forest, one of the core field sites of the National Ecological Observatory Network (NEON) \cite{NEON2019DSM,DataCarpentryRaster}. A DSM is a gridded elevation map that records the elevation of the highest surfaces observed by airborne LiDAR measurements, including tree canopies, buildings, and the ground in unvegetated areas. Observations were collected at a spatial resolution of 1~m and range from approximately 305 to 416~m above sea level. 

This dataset was selected because it exhibits substantial spatial variability and pronounced, irregular transitions between regions of high and low elevation. Such features provide a challenging and informative test case for multi-resolution spatial models, as abrupt boundaries are expected to require finer-scale basis functions to accurately capture localized structure. In contrast, other portions of the study area are relatively homogeneous and can be represented adequately using coarser-scale components. The coexistence of both smooth and highly localized features makes this dataset well-suited for evaluating the ability of multi-resolution models to adaptively allocate model complexity across space.

A simple random sample of $n = 1,200$ locations was drawn from the raster to construct the dataset. The following sections present results from two models fit to these data. The first model employs a two-dimensional extension of Haar wavelet basis, while the second uses a multi-resolution spline basis.

\subsubsection*{Wavelet modeling}
\begin{figure}[H]
    \centering
    \includegraphics[width=\linewidth]{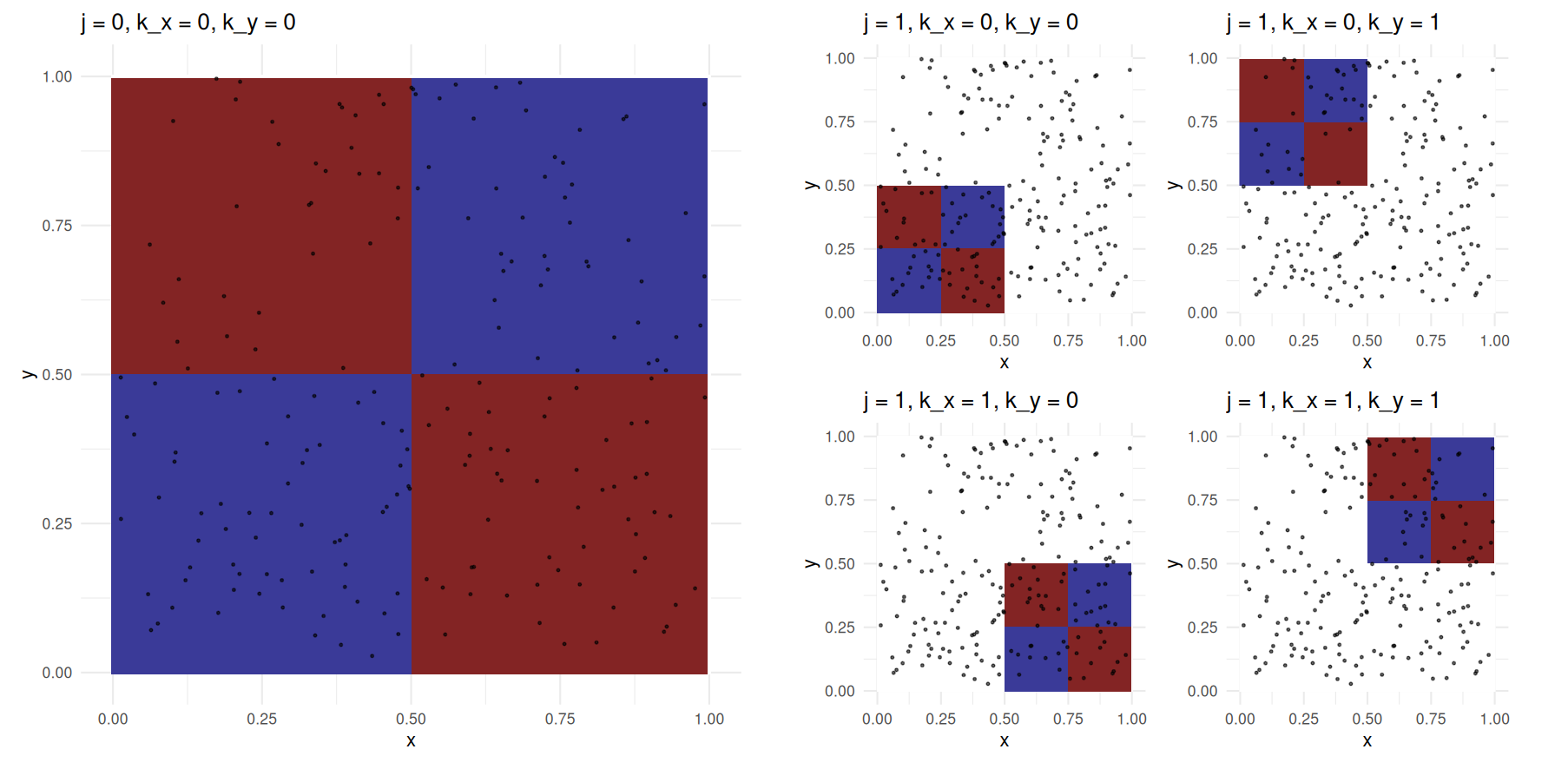}

    \vspace{0.75cm}
    \resizebox{\textwidth}{!}{
\begin{tikzpicture}[
    every node/.style={
        draw,
        rounded corners,
        align=center,
        minimum width=.7cm,
        minimum height=.5cm,
        font=\scriptsize
    },
    arrow/.style={-{Stealth[length=1.8mm]}, thick}
]

\node (b0) at (0,4.8) {Intercept};

\node (psiH000) at (-2.0,3.2) {$\psi^{H}_{0,0,0}$};
\node (psiV000) at ( 0.0,3.2) {$\psi^{V}_{0,0,0}$};
\node (psiD000) at ( 2.0,3.2) {$\psi^{D}_{0,0,0}$};

\node (psiH100) at (-8.25,0) {$\psi^{H}_{1,0,0}$};
\node (psiV100) at (-6.75,0) {$\psi^{V}_{1,0,0}$};
\node (psiD100) at (-5.25,0) {$\psi^{D}_{1,0,0}$};

\node (psiH110) at (-3.75,0) {$\psi^{H}_{1,1,0}$};
\node (psiV110) at (-2.25,0) {$\psi^{V}_{1,1,0}$};
\node (psiD110) at (-0.75,0) {$\psi^{D}_{1,1,0}$};

\node (psiH101) at ( 0.75,0) {$\psi^{H}_{1,0,1}$};
\node (psiV101) at ( 2.25,0) {$\psi^{V}_{1,0,1}$};
\node (psiD101) at ( 3.75,0) {$\psi^{D}_{1,0,1}$};

\node (psiH111) at ( 5.25,0) {$\psi^{H}_{1,1,1}$};
\node (psiV111) at ( 6.75,0) {$\psi^{V}_{1,1,1}$};
\node (psiD111) at ( 8.25,0) {$\psi^{D}_{1,1,1}$};

\draw[arrow] (b0) -- (psiH000);
\draw[arrow] (b0) -- (psiV000);
\draw[arrow] (b0) -- (psiD000);


\foreach \child in {
    psiH100,psiV100,psiD100,
    psiH110,psiV110,psiD110,
    psiH101,psiV101,psiD101,
    psiH111,psiV111,psiD111}
{
    \draw[arrow] (psiH000) -- (\child);
}

\foreach \child in {
    psiH100,psiV100,psiD100,
    psiH110,psiV110,psiD110,
    psiH101,psiV101,psiD101,
    psiH111,psiV111,psiD111}
{
    \draw[arrow] (psiV000) -- (\child);
}

\foreach \child in {
    psiH100,psiV100,psiD100,
    psiH110,psiV110,psiD110,
    psiH101,psiV101,psiD101,
    psiH111,psiV111,psiD111}
{
    \draw[arrow] (psiD000) -- (\child);
}

\end{tikzpicture}
}
    
    \caption{\textbf{Top:} The regions associated with $J=0$ and $J=1$ resolution bases. Each of the red and blue colored regions are described by the three predictors $\left(\psi^H,\,\psi^V,\,\psi^D\right)$, giving a total of 3 resolution-0 predictors and 12 resolution-1 predictors. The simple random sample of spatial locations is plotted as a data cloud. \textbf{Bottom:} Inheritance DAG for 2-dimensional wavelet basis with three orientation types ($H$, $V$, and $D$) and maximum resolution 1. Each predictor at resolution 0 inherits to all predictors at the next resolution since resolution 0 is defined over the entire spatial domain.}
    \label{fig:2DHarrBasis}
\end{figure}
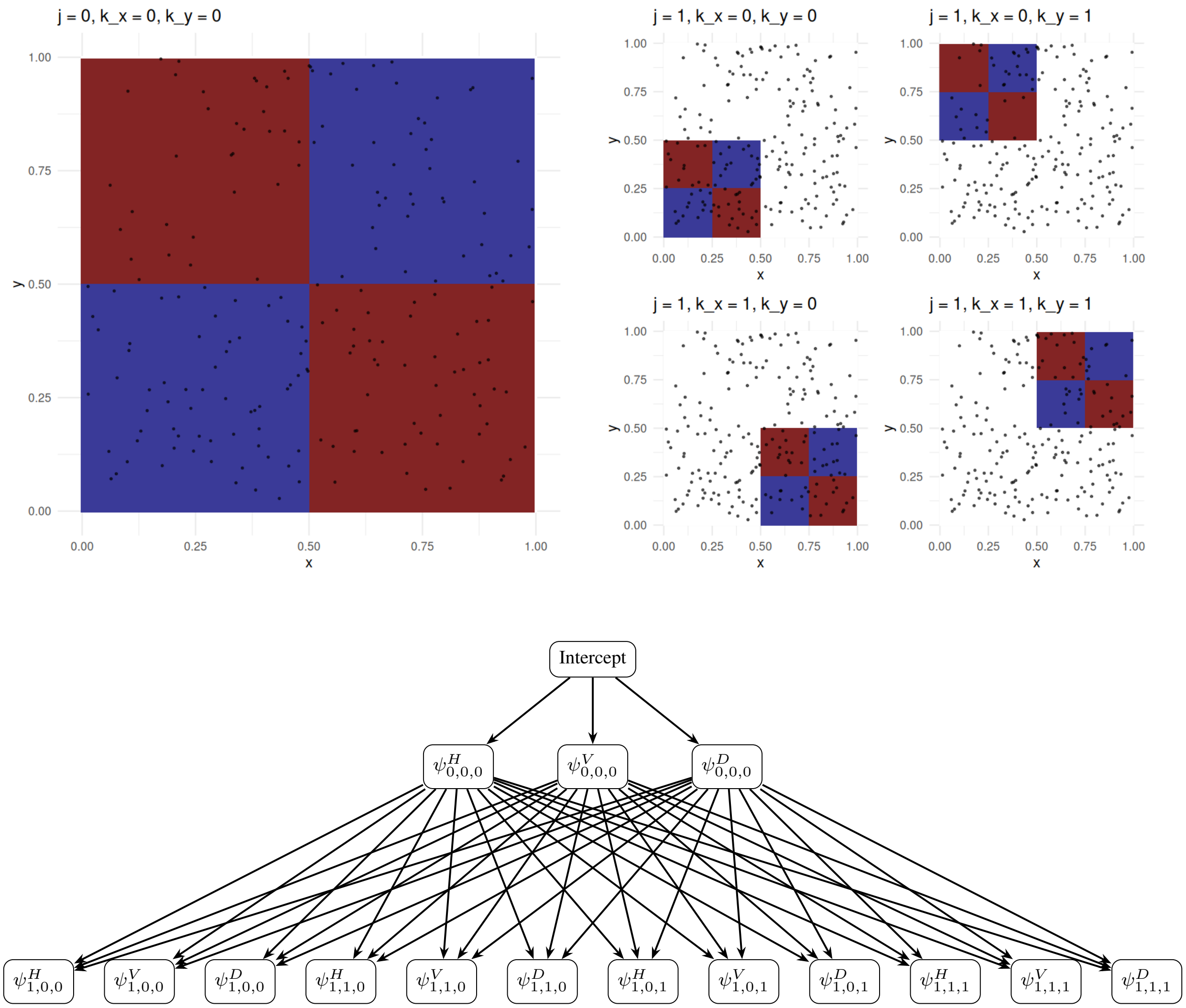

To extend the Haar basis to two dimensions, each region is split into four equal sub-regions, as illustrated in Figure \ref{fig:2DHarrBasis} for resolutions $j=0$ and $j=1$. Instead of modeling each sub-region with one basis as in the one dimensional case, the spatial problem requires the use of three bases with respect to each of these four subregions: top/bottom, left/right, and diagonal. At the $j$th resolution, the sub-region indexed by $(k_x,\,k_y)\in\mathbb Z^+\times\mathbb Z^+$ is modeled using the Haar wavelet predictors $\left(\psi_{j,k_x,k_y}^H,\,\psi_{j,k_x,k_y}^V,\,\psi_{j,k_x,k_y}^D\right)$.

The inheritance structure is determined analogously to the 1D Haar basis, where any predictor of resolution $j+1$ with nonzero values within the domain of a resolution $j$ predictor is considered its child. Formally, parenthood is defined as

$$\mathrm{pa}\left(\psi_{j-1,k_x,k_y}^{a}\right)=\left\{\psi_{j,\kappa_x,\kappa_y}^{\alpha}|A_{j-1}\left(k_x,k_y\right) \subset A_{j}\left(\kappa_x,\kappa_y\right)\right \}\,\,,$$
for orientation types $a,\alpha\in\left\{H,V,D\right\}$, where $A_J(K_x,K_y)$ is the subregion at resolution $j$ indexed by $(K_x,K_y)$. An example of a resulting DAG from this inheritance structure is given in Figure \ref{fig:2DHarrBasis}.

This DAG plot shows that a $J$-resolution 2D Haar wavelet basis results in many spatial predictors, each of which up to resolution $J-1$ have twelve children. Thus, the total number of predictors grows exponentially as shown in Table \ref{tab:numbasis}.

\begin{table}[H]
    \centering
    \begin{tabular}{c|cccccc}
         Maximum Resolution&  0&  1&  2&  3&  4& 5\\
         \hline
         Total 2D Wavelet Predictors&  3&   15&   63&  255& 1023& 4095
         \\
    \end{tabular}
    \caption{Total wavelet bases required for dyadic scaling in two dimensions, not counting the intercept term. Resolution $j$ has $4^j$ sub-regions and thus requires an additional $(3)4^{j}$ bases.}
    \label{tab:numbasis}
\end{table}

A resolution $J=5$ ILS model was fit using an MCMC sample of size 1,000 after discarding 3,000 samples as a burn-in. Posterior medians were used as regression coefficient point-estimates, producing the prediction surface shown in Figure \ref{fig:HarvPreds}.

\subsubsection*{Eigenvalue Ordered Spline (EVOS) Bases for Spatial Data}
\label{s:EVOS}
Consider the multi-resolution spline construction proposed by Tzeng and Huang \cite{Tzeng2018}. Their approach begins with the construction of a thin-plate spline kernel matrix, denoted $\Phi$, using only the spatial coordinates of the observed data locations. Importantly, the basis construction depends solely on geometry, and no response measurements are required at this stage.

To ensure orthogonality regarding low-order polynomial trends and any additional covariates included in the regression model, the thin-plate spline kernel matrix is projected onto the orthogonal complement of the relevant polynomial subspace. Let $Q$ denote the projection matrix onto this orthogonal complement. The projected matrix $Q\Phi Q$ is then eigendecomposed. The resulting eigenvectors form what Tzeng and Huang refer to as a \textit{multi-resolution spline basis}. In the context of ILS, it is helpful to highlight the natural hierarchy in this basis, so it will be called the \textit{eigenvalue ordered spline} (EVOS) basis.

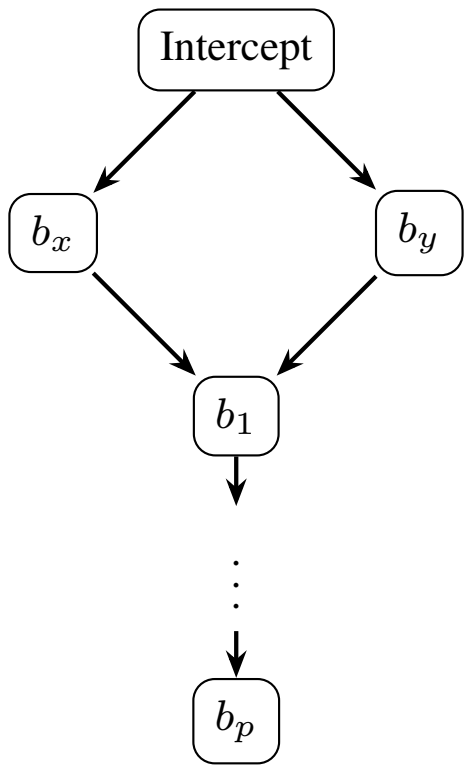
\begin{figure}[H]
\centering
\resizebox{!}{5cm}{
\begin{tikzpicture}[
    every node/.style={
        draw,
        rounded corners,
        align=center,
        minimum width=.5cm,
        minimum height=.5cm,
        font=\scriptsize
    },
    arrow/.style={-{Stealth[length=1.8mm]}, thick}
]

\node (b0) at (0,1.2) {Intercept};

\node (x) at (-1.2,0) {$b_x$};
\node (y) at (1.2,0) {$b_y$};

\node (e1) at (0,-1.2) {$b_{1}$};
\node[draw=none] (dots) at (0,-2.2) {$\vdots$};
\node (ep) at (0,-3.2) {$b_{p}$};

\draw[arrow] (b0) -- (x);
\draw[arrow] (b0) -- (y);

\draw[arrow] (x) -- (e1);
\draw[arrow] (y) -- (e1);

\draw[arrow] (e1) -- (dots);
\draw[arrow] (dots) -- (ep);

\end{tikzpicture}
}
\caption{Inheritance DAG for the EVOS basis in two spatial dimensions using first $p$ eigenvalues. The $b_x$ and $b_y$ terms are the linear effects of the two dimensions.}
\label{fig:EVOS_DAG}
\end{figure}

This construction has several advantages in spatial and spatiotemporal ILS regression. First, the basis functions are orthogonal by construction, both among themselves and with respect to included polynomial covariates. This improves numerical stability and interpretability. Second, the method avoids the need for knot placement, a common source of subjectivity and computational complexity in spline-based modeling. Third, because the eigenvectors are ordered according to their associated eigenvalues, the resulting basis functions are naturally ranked from coarse-scale to fine-scale spatial structure. Consequently, an inheritance graph can be defined according to this resolution hierarchy, enabling direct integration with the ILS framework. Under ILS, finer-scale EVOS predictors are permitted to contribute to the model only if none of their coarser-scale ancestors are heavily shrunk, thereby producing an adaptive, data-driven resolution cutoff.

\begin{figure}[H]
    \centering
    \includegraphics[width=0.485\linewidth]{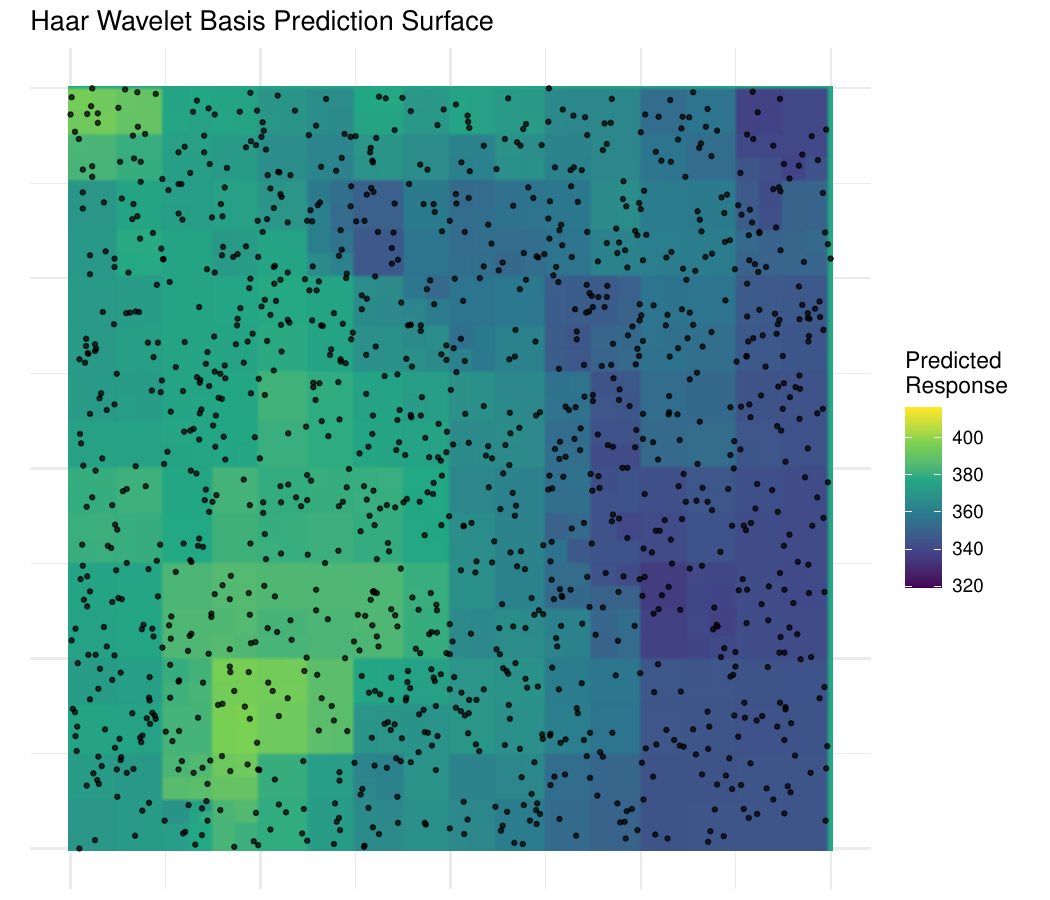}\hfill
    \includegraphics[width=0.485\linewidth]{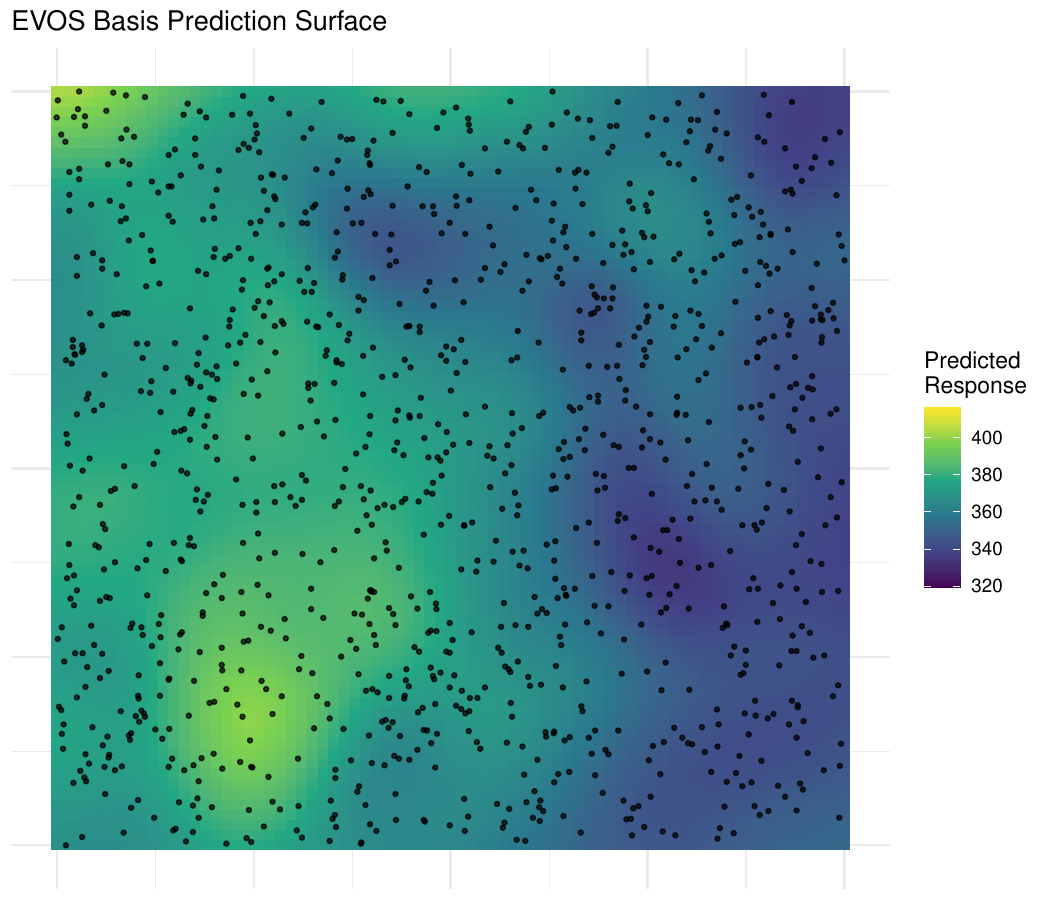}\hfill

    \includegraphics[width=0.485\linewidth]{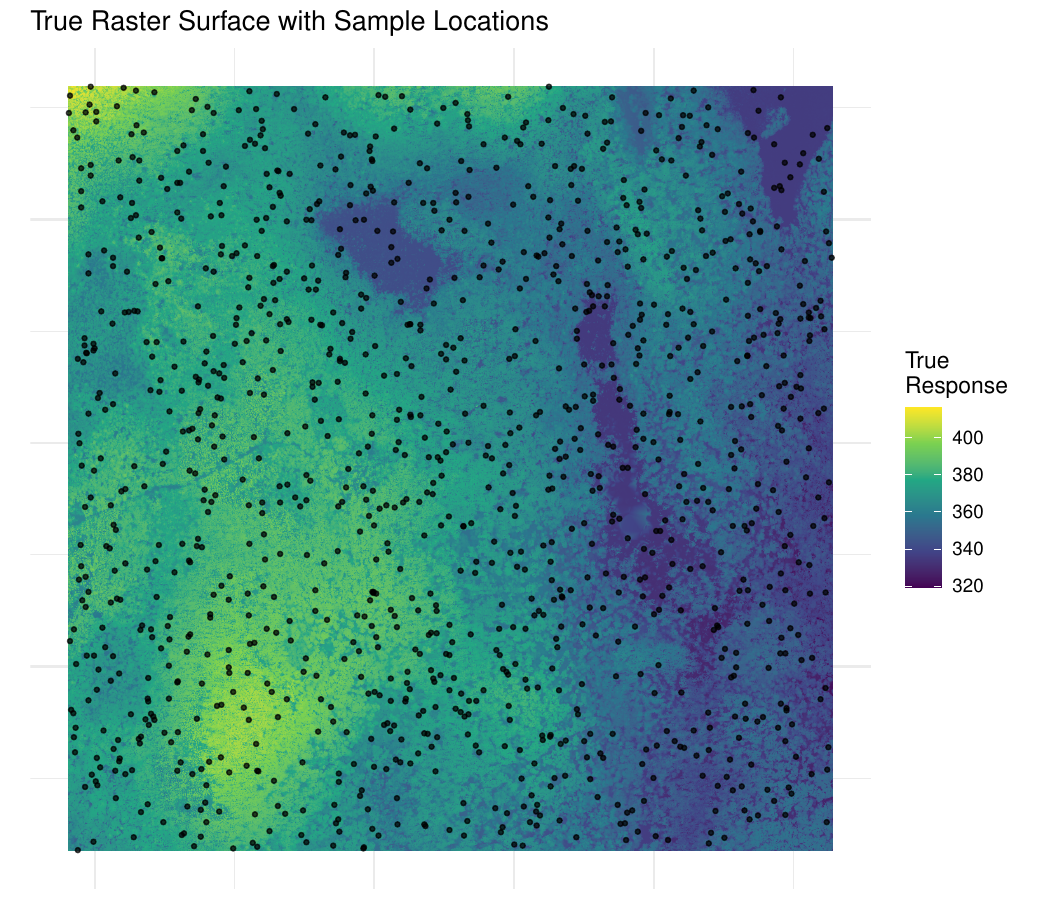}\hfill
    \includegraphics[width=0.485\linewidth]{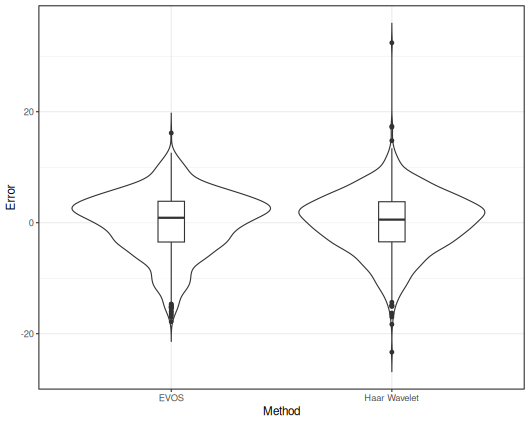}

    \caption{\textbf{Top:} Prediction surfaces for Haar wavelet and EVOS basis ILS models \textbf{Bottom Left:} True surface with simple random sample \textbf{Bottom Right:} Violin plots of prediction error for both models.}
    \label{fig:HarvPreds}
\end{figure}

A resolution $J = 300$ EVOS basis ILS model was fit using 50,000 post-burn-in MCMC samples after discarding the first 2,500 iterations as burn-in. Including the intercept and linear terms, the model contained 303 unshrunk regression coefficients, representing only 7.4\% of the 4,095 coefficients used in the resolution-5 Haar wavelet model. Despite the substantially smaller parameter space and reduced computational burden, the model recovered a remarkable degree of spatial structure, albeit in a smoother form than that obtained with the Haar wavelet basis. As shown in Figure \ref{fig:HarvPreds}, the EVOS and Haar wavelet models had very similar prediction error.

\newpage
Both approaches exhibited strong predictive performance, with only negligible differences in out-of-sample error. The Haar wavelet basis produced slightly lower prediction error overall, outperforming EVOS in approximately 50.4\% of simulation replicates. However, the numerical differences were minimal relative to the scale of the prediction errors. These results indicate that the two methods were effectively indistinguishable from a predictive standpoint. When prediction is the sole objective, either basis appears suitable.

\begin{table}[htbp]
\centering
\begin{tabular}{lrrrrrr}
\hline
Method & Mean & Median & SD & RMSE & MAE \\
\hline
EVOS & -0.00150 & 0.887 & 5.65 & 5.64 & 4.49 \\
Haar Wavelet & -0.000116 & 0.557 & 5.53 & 5.53 & 4.33 \\
\hline
\end{tabular}
\caption{Summary statistics for prediction error by method. SD = standard deviation, RMSE = root mean square error, MAE = mean absolute error.}
\label{tab:prediction_error_summary}
\end{table}

\subsubsection*{Required Resolution Mapping}
An interesting quality of multi-resolution bases is the potential to characterize the spatially varying level of complexity required to adequately represent the fitted surface. Because these bases are constructed as nested sequences of increasingly localized basis functions, each spatial location is associated with a unique chain of descendant basis elements extending from coarse to fine resolution. This structure permits a direct assessment of the finest resolution that contributes meaningfully to the posterior surface at each location.

\begin{figure}[H]
    \centering
    \includegraphics[width=0.45\linewidth]{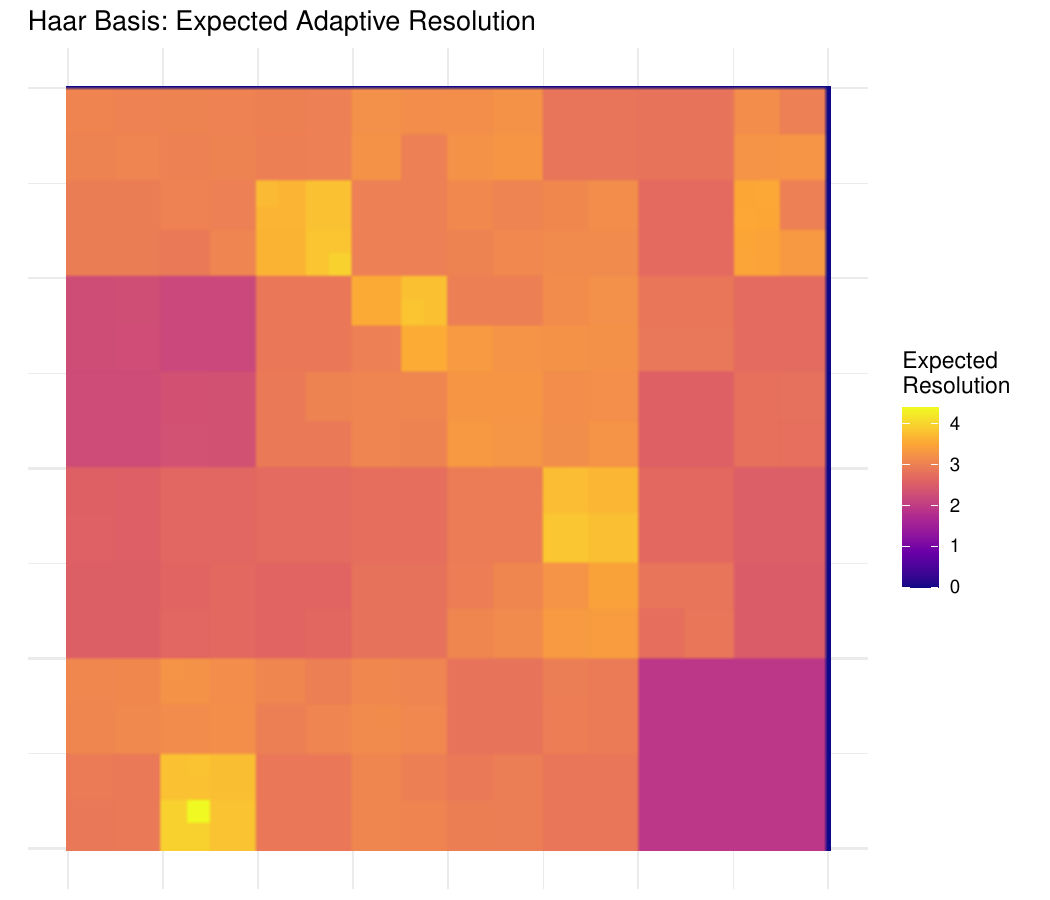}\hfill
    \includegraphics[width=0.45\linewidth]{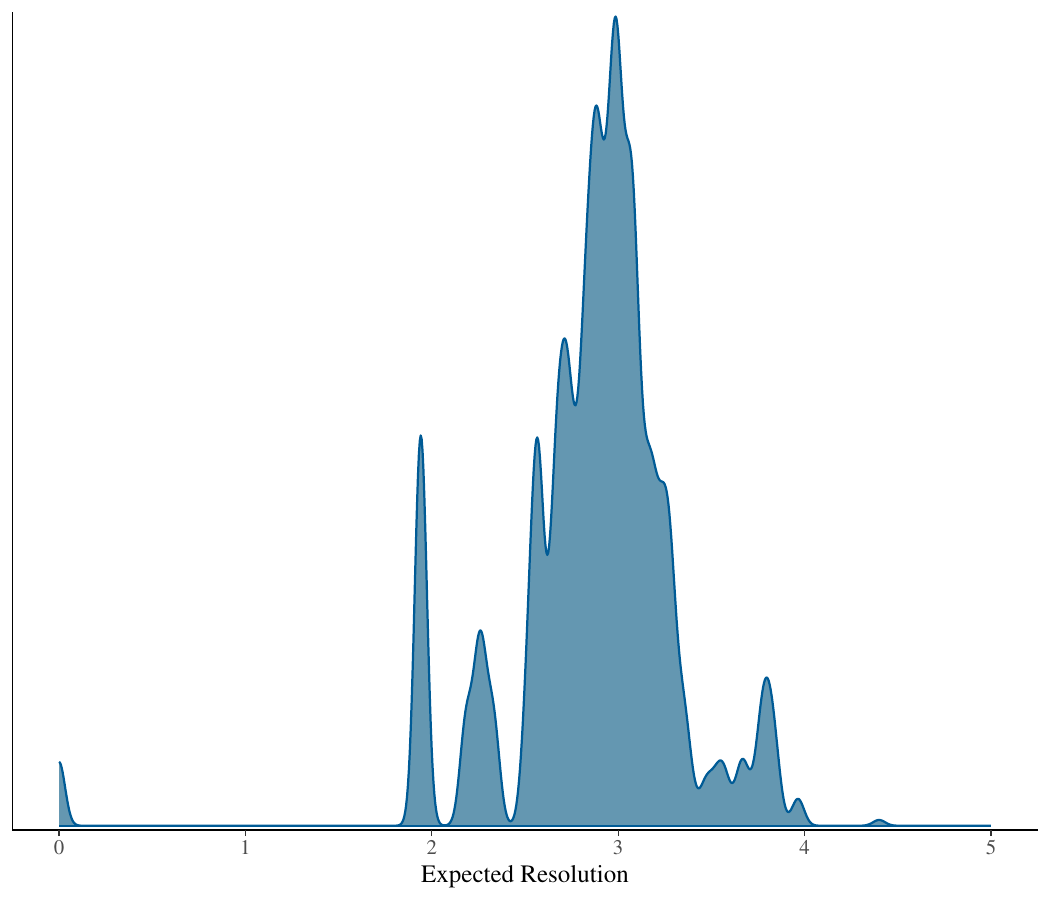}
    
    \vspace{0.5em}

    \includegraphics[width=0.45\linewidth]{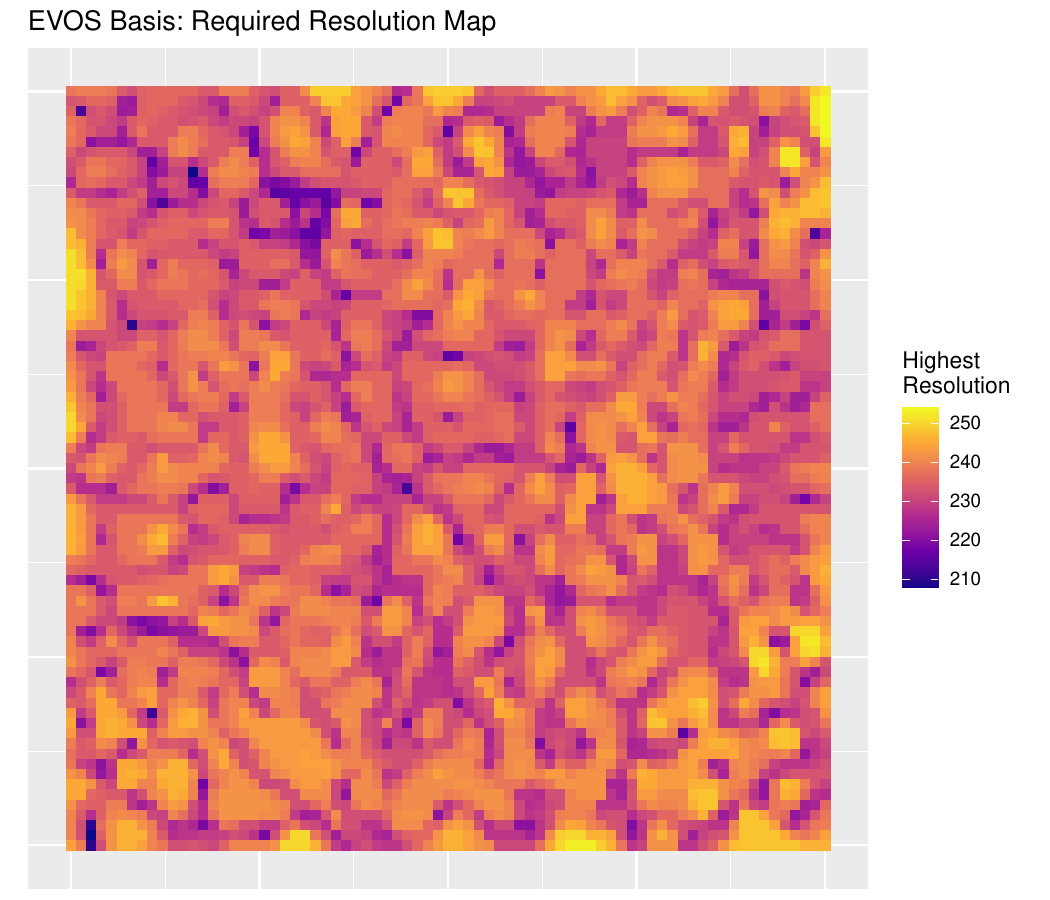}\hfill
    \includegraphics[width=0.45\linewidth]{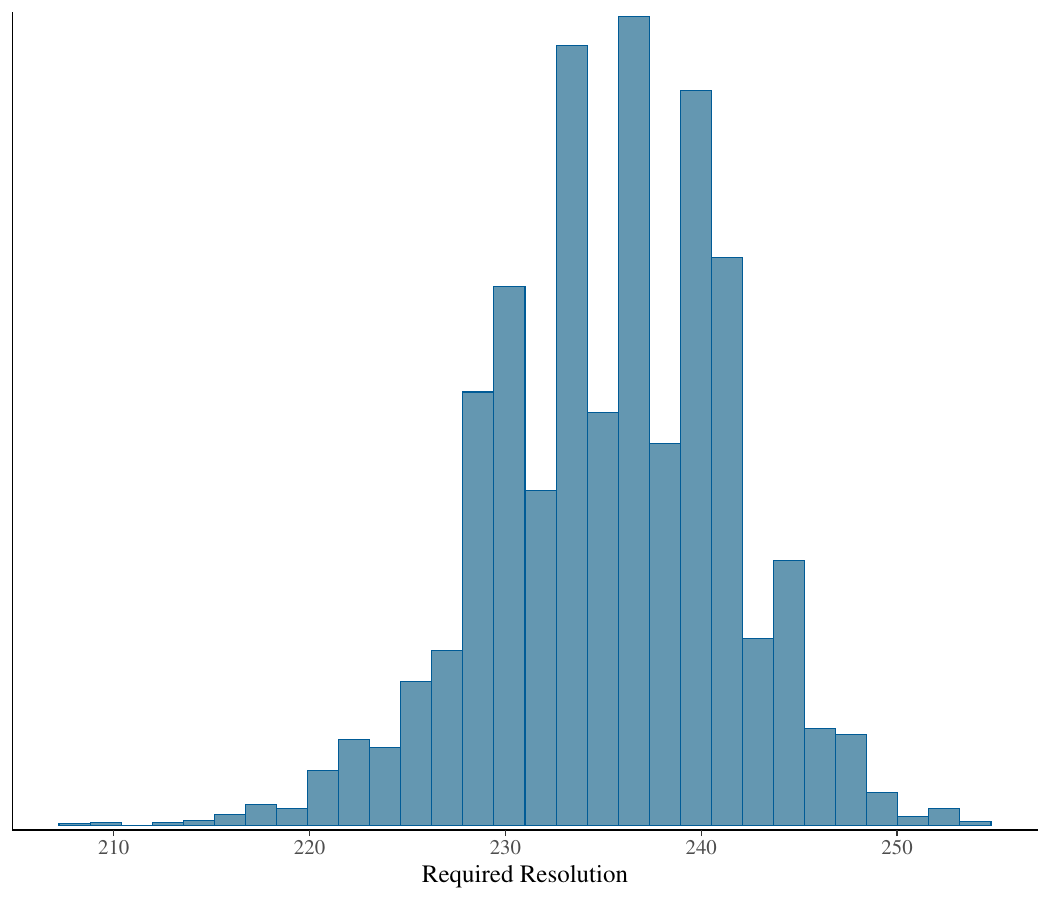}

    \caption{Spatial map and by-pixel distributions of adaptive resolution metrics for Harvard forest models \textbf{Top:} Wavelet basis \textbf{Bottom:} EVOS basis}
    \label{fig:reqRes}
\end{figure}

\paragraph{Wavelet Basis} 
To summarize this behavior for the wavelet basis, we propose a metric termed the \textit{Expected Adaptive Resolution}. At each MCMC iteration, a pseudo-inclusion probability was computed for every regression coefficient as the complement of the shrinkage factor described by \cite{piironen2017hyperprior}:
\[
1 - \kappa_j
=
1 - \frac{1}{1 + n \sigma^{-2} \tau^2 \lambda_j^2},
\]
where $n$, $\sigma$, $\tau$, and $\lambda_j$ are defined in Section \ref{ss:comparableApproaches}. For each cell in the finest-resolution grid, the highest resolution level associated with a predictor having pseudo-inclusion probability greater than 0.5 was recorded. This produced an adaptive resolution surface for a single MCMC draw. Averaging this surface over all posterior draws yielded the expected adaptive resolution surface shown in Figure \ref{fig:reqRes}.

As expected, the highest adaptive resolutions were concentrated along regions with sharp boundaries, where fine-scale modeling was necessary to capture abrupt changes in the response surface. One area required the maximum available resolution of $J = 5$, whereas two relatively homogeneous regions were adequately represented at much coarser resolutions, with expected adaptive resolutions of approximately $J \approx 2$.

\paragraph{EVOS Basis}
Because EVOS basis functions possess global support, each predictor influences the fitted surface over the entire spatial domain. This yields the linear inheritance DAG shown in Figure \ref{fig:EVOS_DAG}. In contrast to the wavelet model, where local basis support allows the required resolution to be determined from the finest-scale unshrunk predictor at each location, such a metric is not meaningful for the EVOS basis. We therefore define required resolution in terms of the additional spatial detail contributed by progressively finer basis functions to the reconstructed surface.

Let
\[
f(\mathbf{s}) = \sum_{j=1}^{J} \beta_j \phi_j(\mathbf{s})
\]
denote the fitted spatial effect, where the basis functions are ordered from coarse to fine resolution. For any truncation level \(r\), define the partial reconstruction

\[
f_r(\mathbf{s}) = \sum_{j=1}^{r} \beta_j \phi_j(\mathbf{s}),
\]
and the remaining refinement $\Delta_r(\mathbf{s}) = f(\mathbf{s}) - f_r(\mathbf{s})$. The quantity \(\Delta_r(\mathbf{s})\) represents the aggregate contribution of all basis functions finer than resolution \(r\).

Rather than introducing an arbitrary tolerance parameter, the required resolution can be defined using posterior credible intervals. Let \(C_{1-p_0}\{\Delta_r(\mathbf{s})\}\) denote the central \((1-p_0)\) credible interval for \(\Delta_r(\mathbf{s})\), where \(p_0\) is a chosen posterior confidence level (e.g., \(p_0 = 0.95\)). The required resolution is then defined as the smallest level \(r\) for which zero lies within this interval $r^*(\mathbf{s})=\min\left\{r :0 \in C_{1-p_0}\{\Delta_r(\mathbf{s})\}\right\}$. Equivalently, \(r^*(\mathbf{s})\) is the coarsest resolution at which the remaining higher-resolution contribution is no longer credibly different from zero at level \(p_0\).

The resulting map of \(r^*(\mathbf{s})\) summarizes the finest level of spatial detail supported by the data at each location. Smooth regions typically require only coarse basis functions, whereas areas exhibiting sharp transitions or localized structure retain finer-resolution terms. This approach avoids the need to specify a practical significance threshold and depends only on the chosen posterior credibility level.

\section{Discussion}
This paper introduced inherited local shrinkage (ILS), a Bayesian regularization framework for adaptive functional form discovery under strong heredity constraints. The method extends the horseshoe prior by replacing independent local shrinkage parameters with structured local scales defined over a directed acyclic graph (DAG). This construction constrains the local shrinkage of increasingly complex terms by the shrinkage assigned to their simpler parent terms.

Across the simulation studies, ILS demonstrated competitive predictive performance relative to both the hierarchical lasso of Bien et al.~\cite{bien2013lasso} and the hierarchical shrinkage priors of Griffin and Brown~\cite{griffin2017hierarchical}. The primary advantage of ILS, however, was its ability to recover underlying functional structure while preserving strong heredity. This behavior was demonstrated in a fourth-degree polynomial simulation, where the method identified the correct hierarchy of main effects, powers, and interactions.

The comparison with existing methods also highlights an important practical distinction. Existing heredity-constrained approaches often require model-specific constraints or variance constructions as hierarchical depth increases, whereas ILS preserves a common prior formulation across arbitrary inheritance DAGs. In practice, extending the model to richer functional representations requires only augmenting the design matrix and specifying the corresponding DAG structure.

The adaptive-resolution applications demonstrated that ILS can also regulate model complexity in multi-resolution basis expansions. Both the Haar wavelet and EVOS bases achieved similar predictive performance while producing substantially different inheritance structures. The Haar basis produced a comparatively wide hierarchy with many basis functions at each resolution level, whereas EVOS generated a much deeper inheritance structure with substantially fewer predictors. Despite using roughly an order of magnitude fewer predictors, the EVOS basis produced predictive accuracy comparable to the wavelet basis while substantially reducing Markov chain Monte Carlo runtime.

Additionally, the adaptive-resolution maps illustrated how ILS can identify where additional spatial detail is supported by the data and where coarse-scale representation is sufficient. Rather than enforcing a uniform resolution across the domain, the method adaptively allocates complexity according to local structure in the response surface. These summaries may also prove useful for informing future spatial sampling strategies. For example, a proxy surface such as biomass or elevation could be combined with an adaptive-resolution map to guide nonhomogeneous sampling designs that allocate additional observations to regions requiring finer spatial detail.

The broader implication is that adaptive functional form discovery and adaptive resolution selection can both be viewed as problems of hierarchical complexity control. By representing candidate predictors as nodes within an inheritance DAG, ILS provides a unified framework for determining where additional model complexity is justified by the data.

Although the present work focused on Gaussian regression with an identity link, the inherited local shrinkage construction is not restricted to this setting. ILS operates through the prior specification on regression coefficients, so it can naturally extend to generalized linear models and other likelihood-based regression frameworks. Extensions to other likelihoods and larger, more computationally demanding inheritance structures remain important directions for future work.

Ecological applications may be particularly well suited to this framework. Modern ecological datasets often contain large collections of environmental predictors observed across multiple spatial and temporal scales, including climate variables, vegetation indices, soil properties, and remotely sensed products. Because ecological relationships are frequently nonlinear and interaction-driven, additive models may fail to capture important structure. Inheritance-based shrinkage offers a principled mechanism for controlling complexity in these settings while preserving interpretability and strong heredity.  

\clearpage
\bibliographystyle{plain}
\bibliography{sources}

\clearpage
\appendix 
\appendix
\begin{center}
    {\Large\bfseries Appendix}
\end{center}

\vspace{1em}

\section{Stochastic Dominance of Shrinkage}

\begin{proposition}[Stochastic dominance of shrinkage under ILS]
\label{prop:ILSdominance}

Let $\mathcal{A}$ denote the index set of polynomial basis functions equipped with the total order $\prec$. Suppose the local shrinkage variances satisfy the inherited local shrinkage condition
\[
\lambda_\alpha^2 \le \lambda_\gamma^2
\quad \text{whenever} \quad
\gamma \prec \alpha\,\,.
\]
Under the global-local shrinkage prior
\[
\beta_\alpha \mid \tau^2, \lambda_\alpha^2 \sim N(0, \sigma^2 \tau^2 \lambda_\alpha^2)\,\,,
\]
the absolute value of the descendant coefficient is stochastically dominated by that of its ancestor. That is, for any $t \ge 0$,
\[
P\!\left(|\beta_\alpha| > t \mid \sigma^2, \tau^2, \lambda_\alpha^2\right)
\le
P\!\left(|\beta_\gamma| > t \mid \sigma^2, \tau^2, \lambda_\gamma^2\right)\,\,,
\]
whenever $\gamma \prec \alpha$.

\end{proposition}

\begin{proof}
Under the conditional prior,
\[
\beta_\alpha \sim N(0,\sigma^2\tau^2 \lambda_\alpha^2)\,\,, 
\qquad
\beta_\gamma \sim N(0,\sigma^2\tau^2 \lambda_\gamma^2)\,\,.
\]
Let $\theta_\alpha^2 = \sigma^2\tau^2 \lambda_\alpha^2$ and $\theta_\gamma^2 = \sigma^2\tau^2 \lambda_\gamma^2$\,\,.  
The inherited local shrinkage condition implies

\[
\theta_\alpha^2 \le \theta_\gamma^2 .
\]
For a zero-mean normal random variable with variance $\theta^2$,
\[
P(|Z| > t) = 2\Phi\!\left(-\frac{t}{\theta}\right)\,\,.
\]
Since $\theta_\alpha \le \theta_\gamma$, we have
\[
\frac{t}{\theta_\alpha} \ge \frac{t}{\theta_\gamma}\,\,.
\]
Because the standard normal CDF $\Phi$ is increasing,
\[
\Phi\!\left(-\frac{t}{\theta_\alpha}\right)
\le
\Phi\!\left(-\frac{t}{\theta_\gamma}\right)\,\,.
\]
Multiplying by 2 yields
\[
P(|\beta_\alpha| > t)
\le
P(|\beta_\gamma| > t)\,\,.
\]
Thus, $|\beta_\alpha|$ is stochastically dominated by $|\beta_\gamma|$, which was to be shown.
\end{proof}
\end{document}